\documentclass[lettersize,journal]{IEEEtran}
\usepackage{amsmath,amsfonts}
\usepackage{algorithmic}
\usepackage{algorithm}
\usepackage{array}
\usepackage[caption=false,font=normalsize,labelfont=sf,textfont=sf]{subfig}
\usepackage{textcomp}
\usepackage{stfloats}
\usepackage{url}
\usepackage{verbatim}
\usepackage{graphicx}
\usepackage{cite}
\usepackage{booktabs} 
\usepackage{tabularx}
\usepackage{multirow}
\usepackage{makecell}

\begin{document}

\title{A Resource-Aware Behavior Reconstruction and Hierarchical Semantic Learning Framework for Host Intrusion Detection}

\author{Youli Tao, Rui Tang, Hao Ren,~\IEEEmembership{Member,~IEEE}, Chengsheng Zhou, Dengzhe Wang, Shuyu Jiang, and Xingshu Chen

\thanks{Youli Tao, Dengzhe Wang are with the School of Cyber Science and Engineering, Sichuan University, Chengdu, Sichuan 610065, China (e-mail: taoyouli@stu.scu.edu.cn; knightwangww@gmail.com.)}
\thanks{Xingshu Chen, Rui Tang, Hao Ren and Shuyu Jiang are with the School of Cyber Science and Engineering, Sichuan University, Chengdu, China and are also with the Key Laboratory of Data Protection and Intelligent Management, Ministry of Education, China (e-mail: chenxsh@scu.edu.cn; tangrscu@scu.edu.cn; hao.ren@scu.edu.cn; shuyu\_jiang@scu.edu.cn).}
\thanks{Chengsheng Zhou is with the School of Cyberspace Security, Beijing University of Posts and Telecommunications, Beijing, 100876, China, and also with the Institute of Security, China Academy of Information and Communications Technology, Beijing, 100191, China (e-mail: 2026080042@bupt.cn).}

\thanks{\textit{Corresponding author: Xingshu Chen, Rui Tang.}}

}



\maketitle

\begin{abstract}
System calls (syscalls) record key interactions between running programs and the operating system kernel, providing fine-grained and minimally intrusive data for host-based intrusion detection systems (HIDS) deployed in cloud and other modern computing environments. 
However, existing methods often model syscalls in their original execution order, where sequences from different processes are interleaved, making informative patterns difficult to extract and raising two questions: whether raw syscall sequences can be reorganized in a way that yields more discriminative representations, and how complex attack patterns can be effectively learned from the reorganized sequences.  
We propose ReSHID, a resource-aware behavior reconstruction and hierarchical semantic learning framework for host intrusion detection. It reconstructs semantically continuous sequences by leveraging syscall semantic invariants to cast subject identity and relationship resolution across PID namespaces as a bipartite matching problem and tracking file descriptor (FD) lifecycles to associate descriptors referring to the same resource. Additionally, features extracted from these sequences are organized into a lightweight subject behavior graph incorporating inter-subject relationships, where GATv2 captures key coordination patterns to model complex attacks involving multiple subjects.
Experimental results show that sequence reconstruction combined with the detection method can improve HIDS performance. Even with a lightweight linear classifier, the proposed method achieves the best results among all compared methods in terms of F1-score (98.64\%), ROC-AUC (99.80\%), and PR-AUC (98.10\%), while reducing the number of $n$-gram features by approximately 75.2\% and 44.1\% compared with the raw sequences and MGFE, respectively.

\end{abstract}


\section{Introduction}
\label{sec:intro}
\IEEEPARstart{H}{ost-based} intrusion detection systems (HIDS) are essential security mechanisms for detecting malicious activities on hosts~\cite{liu2018host}. Attackers can compromise host systems through increasingly diverse attack vectors, including vulnerability exploitation and weak credentials, posing persistent threats to critical services in cloud and other modern computing environments~\cite{afianian2019malware,li2026process}. Regardless of the initial attack vector, achieving an attack objective ultimately requires concrete runtime behaviors to be executed on compromised hosts~\cite{Xiong2022Conan:,han2020unicorn,cheng2024kairos}. Therefore, continuously monitoring and analyzing host runtime activities remains fundamental to identifying malicious behavior and detecting sophisticated attacks.

\begin{figure}[!t]
    \centering
    \includegraphics[width=0.9\columnwidth]{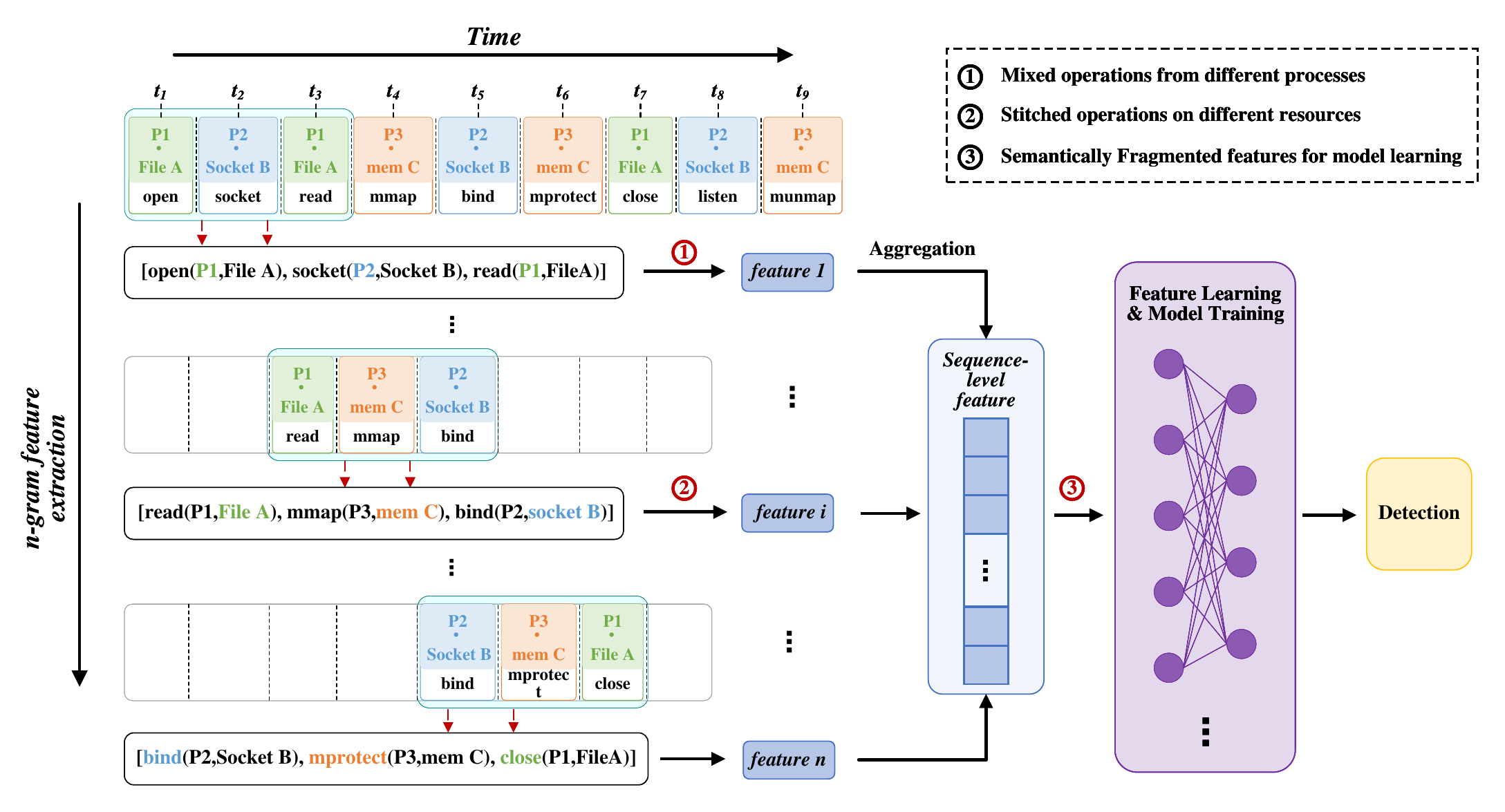} 
    \caption{The raw sequence contains interleaved operations from different subjects on different objects. Directly extracting \(n\)-grams from such a sequence may capture incidental interleavings rather than stable behavioral patterns.}
    \label{fig:introduction_question}
\end{figure}

Such runtime activities can be characterized through various host-level data sources, among which system calls (syscalls) provide a direct view of interactions between user-space programs and the operating system kernel. As the primary interface through which programs request kernel services, syscalls reflect concrete operations on system resources and therefore serve as an important data source for HIDS~\cite{Sworna2023NLP,sekar2024eaudit,zhan2022shrinking}. Accordingly, extensive studies have explored how to characterize host behavior from syscalls for intrusion detection. Existing methods mainly construct behavioral representations from statistical information or syscall sequences. Statistical-feature-based methods characterize program behavior using information such as syscall frequencies~\cite{liao2002using,liu2020statistical}, while sequence-pattern-based methods preserve syscall ordering information to capture finer-grained execution patterns~\cite{warrender1999detecting,zhang2021syscall,zhan2023adversarial}. In particular, the \(n\)-gram model is widely used due to its simplicity and computational efficiency, extracting fixed-length contiguous syscall subsequences through a sliding window to capture local temporal patterns for HIDS~\cite{khreich2017anomaly,karn2020cryptomining,ahmed2020system,subba2021tfidfvectorizer,he2023br,chaudhari2024novel,chew2024real}.

Existing methods typically model host behavior according to the original execution order of syscalls~\cite{creech2013generation,subba2021tfidfvectorizer,zhang2021syscall,Shamim2025Integrating}, implicitly treating temporal proximity as an important indicator of behavioral relevance. Under concurrent multi-process execution, syscalls from multi-process execution can be interleaved by runtime scheduling, such that temporally adjacent operations may originate from different processes and involve unrelated system resources. For clarity, we refer to the task (i.e., process or thread) issuing a syscall as the subject and the resource being operated on as the object. As illustrated in Fig.~\ref{fig:introduction_question}, operations involving different subject-object pairs can be interleaved in the raw syscall sequences. Consequently, semantically related operations on the same object may be separated by unrelated events, whereas unrelated operations may be brought together within the same local window merely due to temporal proximity. Such interleaving can make semantically coherent behaviors appear fragmented in the raw sequence, exhibiting a form of behavioral semantic fragmentation. It causes unrelated operations to be encoded together while separating semantically related ones, thereby introducing incidental patterns into sequence features and weakening the representation of discriminative attack behaviors. This distortion is particularly problematic for complex attacks whose identification often relies on capturing coherent behavioral patterns across multiple operations. Accordingly, two questions are raised: can raw syscall sequences be reorganized according to subject--object relationships to better preserve behavioral semantics and yield more discriminative representations than the original execution order? If so, how can the reconstructed behavioral information be effectively organized and learned to capture complex attack patterns across resources and subjects?

Addressing the first question requires more than simply regrouping syscalls by subject or object. Two practical difficulties must be resolved. First, Linux PID namespaces may assign different identifiers to the same task across namespace views, making it necessary to accurately associate identifiers belonging to the same subject and recover subject relationships with low overhead. Second, file descriptors (FDs) provide only local and dynamically evolving references to underlying system resources~\cite{watson2010capsicum}, and FDs held by different tasks may refer to the same resource. Therefore, reconstructing resource-centric behaviors requires accurate and continuous recovery of cross-task resource associations without intrusive kernel modification. The second question concerns how to effectively exploit the reconstructed behaviors for intrusion detection. Complex attacks may involve coordinated activities across multiple subjects and resources, requiring not only the modeling of local behavioral patterns but also the capture of higher-level dependencies among interacting subjects.

To address these issues, we propose ReSHID, a resource-aware behavior reconstruction and hierarchical semantic learning framework for HIDS. It combines Namespace-aware Subject Identity and Relation Association via Invariant Constraints (NIRA) with Object Identity Resolution via FD Propagation (FDProp) to reconstruct syscall sequences according to subject--object relationships, and employs Hierarchical Behavioral Semantic Learning method (HBSL) for end-to-end learning of key behavioral patterns for intrusion detection. The main contributions of this paper are as follows:

\begin{itemize}

\item First, we systematically analyze the semantic fragmentation of syscall sequences caused by concurrent multi-process scheduling and propose a sequence reconstruction method based on NIRA and FDProp. 
NIRA formulates subject relation recovery as a bipartite matching problem and exploits constraints derived from syscall semantic invariants to associate identifiers of the same subject across namespace views.
FDProp tracks the FD lifecycle according to syscall semantics to continuously recover mappings between FDs and underlying resources across tasks. By recovering these associations, the method reorganizes interleaved syscalls into more semantically coherent behavioral sequences, providing a more informative basis for extracting intrusion-relevant patterns. 

\item Second, we propose HBSL, which extracts object-level operation features from the reconstructed sequences and aggregates them by associated subject to obtain subject-level representations. Rather than directly modeling syscalls as a flat sequence or graph, HBSL organizes these representations into a lightweight subject behavior graph, where subject relationships explicitly encode the structural dependencies among runtime behaviors. It then applies GATv2 to adaptively model the behavioral dependencies among interacting subjects and aggregate informative neighborhood semantics, enabling hierarchical learning from resource-level operations to cross-subject malicious coordination patterns.

\item Finally, given that mainstream HIDS datasets provide limited syscall information and are less representative of contemporary attacks, we construct and publicly release a new dataset\footnote{The dataset is available at: https://github.com/TYLkhjy/A-Linux-HIDS-dataset.git.}, covering diverse contemporary benign and attack scenarios while preserving fine-grained syscall information such as key arguments.Experiments show that sequence reconstruction can improve HIDS performance compared with raw syscall sequences, whose interleaved structure makes discriminative patterns of complex attacks more difficult to learn. Even with a lightweight linear classifier, our method outperforms representative HIDS methods, achieving an F1-score of 98.64\%, a ROC-AUC of 99.80\%, and a PR-AUC of 98.10\%.

\end{itemize}

\section{Related Work}
Syscall-based HIDS studies can be broadly categorized into three groups: statistical features, sequence patterns, and semantic characteristics. In addition, provenance-based intrusion detection systems (PIDSs) represent another line of research, organizing the processes, files, network entities, and their dependencies involved in syscall events into provenance graphs for intrusion detection~\cite{han2020unicorn,han2021sigl,wang2022threatrace,cheng2024kairos}. In contrast to PIDSs, our work focuses on whether reconstructing syscall sequences before feature extraction improves the discriminative capability of behavioral representations, and on how effective detection features can be further learned from these sequences for HIDS. 
Thus, we discuss related work in these areas.

\subsection{HIDS Based on Statistical Features}

Statistical-feature-based HIDS transforms variable-length syscall traces into fixed-length representations using information such as syscall counts and frequency distributions, followed by machine learning or deep learning for detection. Early studies characterized program behavior using syscall frequencies, minimum cross-entropy, and statistical measures~\cite{liao2002using,yeung2003host,joraviya2024ab}, while \cite{chen2005application} further compared frequency-based and TF--IDF encodings and showed that different statistical representations affect downstream detection performance. Subsequent studies incorporated richer statistical properties, such as transition information~\cite{wang2006profiling} and occurrence-frequency patterns of \(n\)-grams~\cite{liu2020statistical}. Overall, these methods offer relatively simple and efficient behavioral representations, but they do not explicitly preserve syscall ordering, which is important for a more comprehensive understanding of host behavior.

\subsection{HIDS Based on Sequence Patterns}

To preserve syscall ordering information, sequence-pattern-based HIDS models syscall traces as ordered sequences and detects anomalies by learning temporal dependencies, state transitions, or local behavioral patterns. Early studies established this paradigm by modeling normal UNIX process behavior using syscall sequences~\cite{forrest1996sense,warrender1999detecting,hofmeyr1998intrusion,sekar2000fast}. Subsequent work extended this paradigm through non-contiguous syscall patterns~\cite{creech2013semantic} and deep generative sequence models for capturing long-term dependencies~\cite{wang2022unsupervised}. Among these approaches, \(n\)-gram models have been widely used in applications such as ransomware and cryptomining detection~\cite{karn2020cryptomining,ahmed2020system,chaudhari2024novel,chew2024real}, including variable-length \(n\)-grams combined with one-class SVMs~\cite{khreich2017anomaly}. To address the high dimensionality and sparsity of \(n\)-gram features, later studies introduced TF--IDF weighting and truncated SVD~\cite{subba2021tfidfvectorizer}, as well as multi-granularity feature compression methods~\cite{he2023br}. However, reducing the feature space may also discard fine-grained behavioral information.

\subsection{HIDS Based on Semantic Characteristics}

Researchers have explored various approaches to enrich the semantic representation of syscall sequences. Some methods enhance sequence semantics by modeling contextual and structural relationships, including predictive sequence modeling~\cite{ring2021methods}, syscall abstraction and differential encoding~\cite{zhang2021syscall}, critical behavioral units and their contextual dependencies~\cite{zhan2023adversarial}, and image-based representations of syscalls and associated metadata~\cite{joraviya2024dl}. Others incorporate syscall arguments and execution context, such as argument-specific anomaly modeling~\cite{Mutz2006Anomalous}, clustering and Markov-based argument analysis~\cite{maggi2008detecting}, graph-based contextual anomaly modeling with syscall and file-path information~\cite{el2022contextualizing}, and system-call graphs constructed from syscall categories and subcategories~\cite{araujo2025enhancing}. Overall, these methods enrich syscall semantics through contextual modeling or additional runtime information, but mainly use such information for feature encoding or anomaly statistics, without extending its use beyond feature-level modeling.

\section{Preliminaries and Problem}
In this section, we define the terminology used throughout this work and describe the impact of semantic fragmentation as well as the general idea for addressing it.

\subsection{Definitions}
\label{subsec:definitions}

\textit{Definition 1 (Raw Syscall Sequence):}
For a target program execution, the generated $L$ syscall operations are ordered by timestamp to form the corresponding raw syscall sequence:
\begin{equation}
\mathcal{O} = \langle \mathit{op}_1, \mathit{op}_2, \dots, \mathit{op}_L \rangle 
\end{equation}
$\mathit{op}_i = (t_i, s_i, \mathit{subj}_i, f_i, \mathit{obj}_i)$, where $t_i$ denotes the timestamp of the operation, $s_i$ denotes the namespace of the subject at that time, $\mathit{subj}_i$ denotes the process or thread that issues the syscall, referred to as the subject, $f_i$ denotes the syscall type, and $\mathit{obj}_i$ denotes the system resource accessed by the subject, such as a file, pipe, or socket, referred to as the object. The operations satisfy $t_{i-1} \leq t_i$ for $2 \leq i \leq L$.

\textit{Definition 2 (Feature Density):}
Let $\mathcal{F}_{\mathrm{raw}}^{(n)}$ denote the set of $n$-gram features constructed from the raw syscall sequences in the training data. For a given sequence $\mathcal{O}$, its feature vector is denoted by $\mathbf{x}_{\mathrm{raw}} = \Phi_{\mathrm{raw}}^{(n)}(\mathcal{O})$. We define its feature density as:
\begin{equation}
\rho_{\mathrm{raw}} = \frac{\|\mathbf{x}_{\mathrm{raw}}\|_0}{|\mathcal{F}_{\mathrm{raw}}^{(n)}|} 
\end{equation}
where $\|\mathbf{x}_{\mathrm{raw}}\|_0$ denotes the number of nonzero elements in $\mathbf{x}_{\mathrm{raw}}$, and $|\mathcal{F}_{\mathrm{raw}}^{(n)}|$ denotes the dimensionality of the feature space. A smaller $\rho_{\mathrm{raw}}$ indicates that the representation of an individual sample is sparser in the feature space.

\textit{Definition 3 (HIDS Objective):}
Given a dataset containing $N$ execution samples, $\mathcal{D} = \{(\mathcal{O}_i, y_i)\}_{i=1}^{N}$, where $\mathcal{O}_i$ denotes the raw syscall sequences of the $i$-th sample and $y_i \in \{0, 1\}$ denotes its benign or malicious label, a syscall-based HIDS aims to extract behavioral features from syscall sequences and use them to identify anomalous behavior. The detection process can be expressed as:
\begin{equation}
\hat{y}_i = h\left( \Phi(\mathcal{O}_i); \theta \right) 
\end{equation}
where $\Phi(\cdot)$ denotes the mapping from syscall sequences to sample-level behavioral features, $h(\cdot;\theta)$ denotes the detection model parameterized by $\theta$, and $\hat{y}_i \in \{0, 1\}$ denotes the predicted label.

\subsection{Threat Model}
\label{subsec:threat_model}
We consider that the adversary gains the ability to execute user-space code on a target Linux host through vulnerable applications, exposed network services, or similar attack vectors, and can further carry out malicious activities such as privilege escalation. In this paper, we focus on distinguishing benign from malicious behavior based on the syscalls generated during application execution.

The operating system kernel and syscall collection mechanism are assumed to be trusted and capable of reliably recording syscalls together with the necessary execution context. Attacks that tamper with the kernel, disrupt syscall collection, or otherwise compromise the integrity of collected data are beyond the scope of this paper.

\subsection{Problems and Defense Strategies}
\label{subsec:semantic_fragmentation}

\subsubsection{Impact of Semantic Fragmentation}

As we discuss in Section~\ref{sec:intro}, such interleaving can make semantically coherent behaviors appear fragmented in the raw sequence, exhibiting a form of behavioral semantic fragmentation. Let the feature representation of the raw syscall sequence $\mathcal{O}$ be denoted by $\mathbf{x}_{\mathrm{raw}}$. We divide its features into effective features $\mathbf{x}_{\mathrm{eff}}$, which can consistently reflect differences between classes, and unstable features $\mathbf{x}_{\mathrm{frag}}$, which arise from concurrent interleaving. Let their numbers be $s_{\mathrm{eff}}$ and $s_{\mathrm{frag}}$, respectively. The proportion of effective discriminative information is then defined as $\eta_{\mathrm{raw}} = s_{\mathrm{eff}}/(s_{\mathrm{eff}}+s_{\mathrm{frag}})$.
As $s_{\mathrm{frag}}$ increases, $\eta_{\mathrm{raw}}$ decreases accordingly. To characterize the impact of $\mathbf{x}_{\mathrm{frag}}$ on discriminability, let $\bar{q}_{\mathrm{frag}}$ denote the average perturbation introduced by each unstable feature. The total perturbation can then be expressed as $Q_{\mathrm{frag}} = s_{\mathrm{frag}}\bar{q}_{\mathrm{frag}} = s_{\mathrm{eff}}(1-\eta_{\mathrm{raw}})\bar{q}_{\mathrm{frag}}/\eta_{\mathrm{raw}}$.
Following the measurement principle of Fisher discriminant analysis, let $\Delta$ denote the average discriminative difference between benign and malicious samples, and let $\sigma_{\mathrm{eff}}^2$ denote the within-class dispersion produced by the effective behaviors. We then define the following discriminability measure:
\begin{equation}
J_{\mathrm{raw}} = \frac{\Delta^2}{\sigma_{\mathrm{eff}}^2 + s_{\mathrm{eff}} \dfrac{1-\eta_{\mathrm{raw}}}{\eta_{\mathrm{raw}}} \bar{q}_{\mathrm{frag}}} 
\end{equation}

When all other factors remain unchanged, a lower $\eta_{\mathrm{raw}}$ indicates greater perturbation introduced by unstable features and therefore a lower $J_{\mathrm{raw}}$. We further analyze its impact on feature-space sparsity and computational efficiency. 
Taking $n$-grams as an example, $|\mathcal{F}_{\mathrm{raw}}^{(n)}|$ denotes the feature dimensionality observed in the training data, while $\|\mathbf{x}_{\mathrm{raw}}\|_0$ denotes the number of nonzero features in an individual sample. Since a sequence of length $L$ can generate at most $L-n+1$ contiguous $n$-grams, we have $\|\mathbf{x}_{\mathrm{raw}}\|_0 \le L-n+1$. According to the feature density defined above, $\rho_{\mathrm{raw}} = \|\mathbf{x}_{\mathrm{raw}}\|_0 / |\mathcal{F}_{\mathrm{raw}}^{(n)}|$. For a fixed sequence, $\|\mathbf{x}_{\mathrm{raw}}\|_0$ remains unchanged. Therefore, when semantic fragmentation introduces additional unstable features and increases $|\mathcal{F}_{\mathrm{raw}}^{(n)}|$, $\rho_{\mathrm{raw}}$ decreases, resulting in a higher-dimensional and sparser feature representation and reducing the efficiency of model training and detection.

\subsubsection{Defense Strategies}
Our defense strategy consists of two stages: reorganizing raw sequences around underlying objects to mitigate semantic fragmentation and hierarchically learning key anomalous patterns from these sequences. Let the set of underlying resources involved in  $\mathcal{O}$ be $\mathcal{B}=\{o_1, o_2, \dots, o_K\}$. $\mathcal{O}$ is reconstructed as $\mathcal{S}_{\mathrm{obj}} = \left\{ \mathcal{O}_{o_1}, \mathcal{O}_{o_2}, \dots, \mathcal{O}_{o_K} \right\}$, where each $\mathcal{O}_{o_k}$ contains only operations on the same object $o_k$ while preserving their relative execution order. Let $\Psi$ denote the raw syscall vocabulary, with $|\Psi|=S$. Suppose the objects in $\mathcal{B}$ belong to $C$ object categories, where $C \leq K$, and let $\Psi_i\subseteq\Psi$ denote the syscall vocabulary associated with the $i$-th object category, with $|\Psi_i|=S_i$. Before reconstruction, the theoretical candidate space of $n$-grams is $|\Omega_{\mathrm{raw}}^{(n)}|=S^n$, whereas after reconstruction it satisfies $|\Omega_{\mathrm{obj}}^{(n)}| \leq \sum_{i=1}^{C} S_i^n$. Therefore,
\begin{equation}
\frac{|\Omega_{\mathrm{obj}}^{(n)}|}{|\Omega_{\mathrm{raw}}^{(n)}|} \leq \sum_{i=1}^{C} \left(\frac{S_i}{S}\right)^n \leq C \left(\frac{S_{\max}}{S}\right)^n 
\end{equation}
where  $S_{\max}=\max_{1\leq i\leq C} S_i \leq S$. This result shows that object-level semantic constraints can limit the candidate combinations of operations, helping to reduce unstable features caused by the interleaving of operations on different objects and increase the information density of the resulting features.

After reconstruction, an $n$-gram model is used to extract the object-level operation feature of $o_k$, denoted as $\mathbf{z}_{o_k} = \phi(\mathcal{O}_{o_k})$. For a subject $u$, let $\mathcal{B}(u)$ denote the set of objects operated on by that subject. These object-level features are aggregated as $\mathbf{h}_u=\operatorname{AGG}(\{\mathbf{z}_{o_k}\mid o_k\in\mathcal{B}(u)\})$. We then construct a lightweight subject behavior graph $G$, where $\mathbf{h}_u$ is used as the feature of node $u$, and subject relationships are used as edges to characterize runtime dependencies among subjects. 
GATv2 further propagates and aggregates behavioral information among related subjects to capture coordinated behavior patterns, and the resulting node representations are pooled as $\Phi(\mathcal{O})=\operatorname{POOL}(H^{(L)})$ to obtain the sample-level behavioral representation for HIDS.

\section{Proposed Method}
\begin{figure*}[t]
    \centering
    \includegraphics[width=0.9\textwidth]{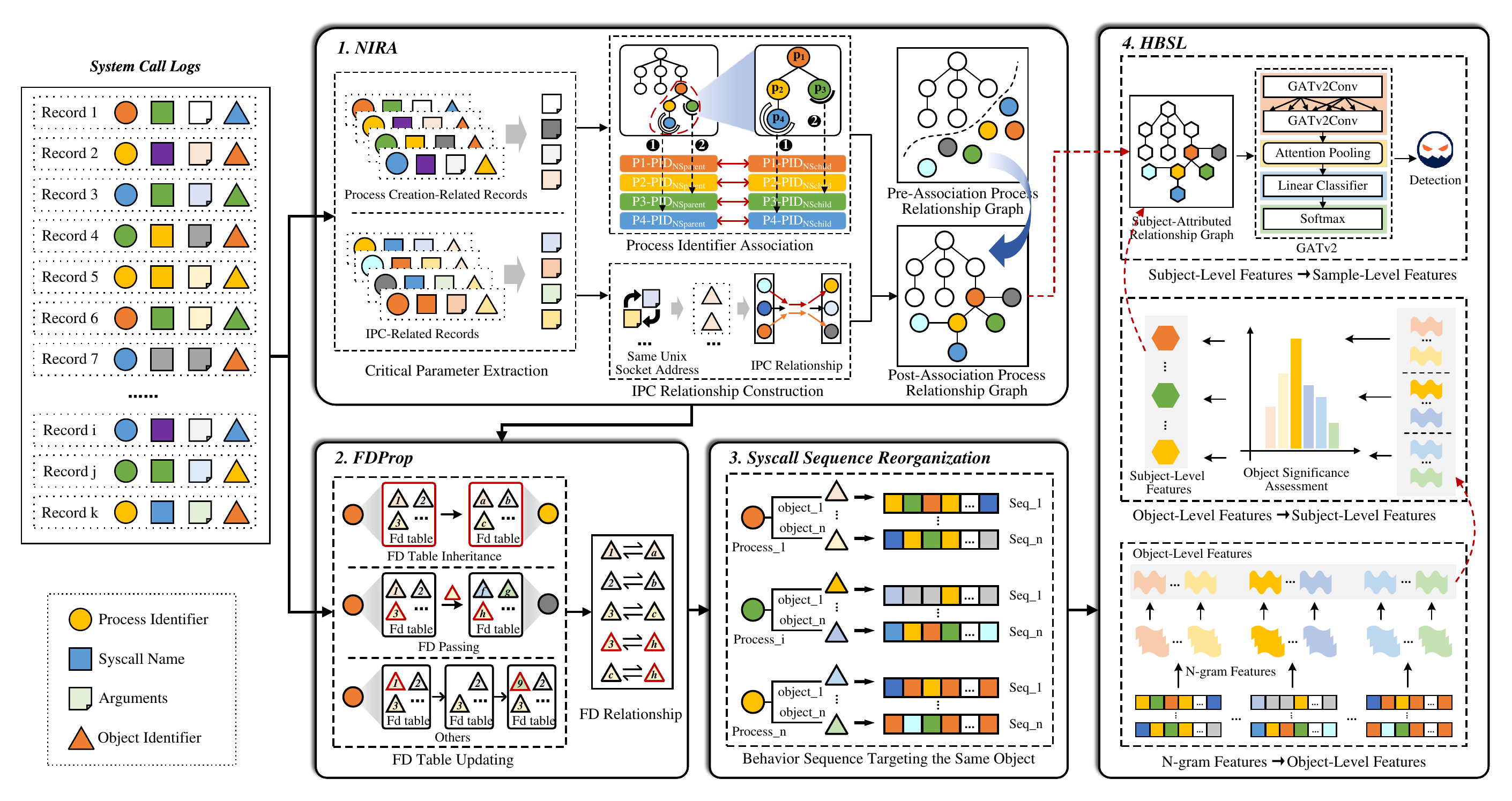}
    \caption{Overview of ReSHID, illustrating the key phases from raw syscall collection to semantic sequence reconstruction and detection.}
    \label{fig:research_framework}
\end{figure*}

In this section, we describe the proposed method in detail. As illustrated in Fig.~\ref{fig:research_framework}, the method consists of four main steps. First, to support sequence reconstruction, NIRA associates subject identifiers that refer to the same task across different PID namespaces and recovers subject relationships. Second, FDProp associates FDs that refer to the same underlying system resource dynamically. Third, based on these associations, the raw sequences are reconstructed into multiple new sequences, each centered on a specific object. Finally, HBSL extracts subject-level behavioral representations from these sequences to construct a lightweight subject behavior graph and builds an end-to-end detection model to capture key malicious coordination patterns.

\subsection{NIRA}
\subsubsection{Task Creation Relation Resolution}
Using Sysdig, a widely used syscall collection tool, as an example, a successful task creation operation typically produces two corresponding records: one from the parent task that performs the creation operation and the other from the newly created child task. The collected syscall records are uniformly abstracted as:
\begin{equation}
e = \langle \tau, ns, id_g, tgid_g, p\_tgid_g, id_v, tgid_v, \mathcal{F}, res \rangle 
\end{equation}
$\tau$ and $ns$ denote the event timestamp and the PID namespace in which the recorded task resides, respectively. $id_g$ and $tgid_g$ denote the global task identifier and thread-group identifier from the collection perspective, respectively, while $p\_tgid_g$ denotes the parent task identifier of the main thread identified by $tgid_g$. $id_v$ and $tgid_v$ denote the corresponding local identifiers in the current PID namespace. $\mathcal{F}$ denotes the syscall-specific arguments, and $res$ denotes the syscall return value. For a parent-side creation record $e_p$, $e_p.res > 0$, and the return value represents the local identifier of the newly created task in the parent task's PID namespace view. For a child-side creation record $e_c$, $e_c.res = 0$, indicating that the child task has been created successfully. However, $e_p.res$ does not necessarily equal the identifier of the created child task from the global PID namespace view. Therefore, task creation relationships cannot be directly recovered from the raw records.

A syscall record can be described along five dimensions: time, space, subject, operation, and object. For the same task creation operation, $e_p$ and $e_c$ should satisfy corresponding relationships across these dimensions. Since the child task acts as the object being created, its identifier is precisely what needs to be determined through association. Accordingly, NIRA exploits complementary information observable from the spatial, subject, operation, and temporal dimensions and formulates creation-relation recovery as a bipartite matching problem under consistency constraints. Given the parent-side record set $E_p$ and the child-side record set $E_c$, all potential record correspondences form the Cartesian product $\mathcal{U} = E_p \times E_c$ when no constraints are applied. NIRA constructs a set of consistency constraints and defines the record pairs satisfying all of them as the feasible relation set:
{\small
\begin{equation}
\mathcal{R} = \left\{ (e_p, e_c) \in \mathcal{U}\middle|\; C_{\mathrm{ns}} \land C_{\mathrm{group}} \land C_{\mathrm{parent}} \land C_{\mathrm{flag}} \land C_{\mathrm{time}} \right\} 
\end{equation}
}
$C_{\mathrm{ns}}$ corresponds to the spatial dimension, $C_{\mathrm{group}} \land C_{\mathrm{parent}}$ to the subject dimension, $C_{\mathrm{flag}}$ to the operation dimension, and $C_{\mathrm{time}}$ to the temporal dimension. 
In this way, the initial candidate relation space $\mathcal{U}$ is reduced to $\mathcal{R} \subseteq \mathcal{U}$. These dimensional constraints are derived from the system semantics of task creation operations, as summarized in Table~\ref{tab:nira_constraints}.

The spatial constraint further needs to account for dynamic changes in PID namespaces. The PID namespace of a newly created task depends not only on the namespace of the creating subject but also on the namespace context used for subsequent child creation. To capture this behavior, NIRA maintains a dual-state namespace context for each task $u$:
\begin{equation}
\mathcal{S}(u) = \langle \mathcal{N}_{\mathrm{cur}}(u), \mathcal{N}_{\mathrm{child}}(u) \rangle 
\end{equation}
where $\mathcal{N}_{\mathrm{cur}}(u)$ denotes the current PID namespace of the subject, and $\mathcal{N}_{\mathrm{child}}(u)$ denotes the target PID namespace for subsequent child creation. For example, \texttt{unshare(CLONE\_NEWPID)} updates $\mathcal{N}_{\mathrm{child}}(u)$, whereas \texttt{clone(CLONE\_NEWPID)} affects only the child process created by that particular operation. NIRA derives the PID namespace in which a candidate child task should reside from this namespace context and the semantics of the current creation operation, and then uses the result to determine spatial consistency.

\begin{table*}[t] 
\centering
\caption{Multi-dimensional constraints for subject creation relation recovery}
\label{tab:nira_constraints}
\begin{tabularx}{\textwidth}{l l X} 
\toprule
\textbf{Dimension} & \textbf{Constraint} & \textbf{Criterion} \\
\midrule
Spatial & $C_{\mathrm{ns}}$ & Determine whether the candidate child task resides in the expected PID namespace based on the namespace context $\mathcal{S}(u)$ of the creating subject and the semantics of the creation operation. \\[3pt]
Subject & $C_{\mathrm{parent}} \land C_{\mathrm{group}}$ & Determine whether the subject relationship is consistent by jointly considering the parent-subject information and thread-group relationship of the candidate child record. For example, when $\mathcal{F}$ contains \texttt{CLONE\_THREAD}, the condition $e_c.tgid_g = e_p.tgid_g$ must hold. \\[3pt]
Operation & $C_{\mathrm{flag}}$ & Determine whether the parent-side and child-side records have compatible creation semantics according to the flags in $\mathcal{F}$ related to thread groups, resource sharing, and namespace creation. \\[3pt]
Temporal & $C_{\mathrm{time}}$ & Require the parent-side and child-side records to satisfy $|e_c.\tau - e_p.\tau| \leq \Delta\tau$. \\
\bottomrule
\end{tabularx}
\end{table*}

However, under concurrent task creation, a single record may still match multiple candidate records. To resolve this ambiguity, NIRA treats $E_p$ and $E_c$ as the two node sets of a bipartite graph, and uses the feasible record pairs in $\mathcal{R}$ as edges to construct $G_B = (E_p, E_c, \mathcal{R})$. For each candidate edge $(e_p, e_c) \in \mathcal{R}$, the matching cost is defined by the temporal distance: $w(e_p, e_c) = |e_p.\tau - e_c.\tau|$.
Since each successful subject creation operation corresponds to a unique parent--child record pair, the recovered relations should satisfy a one-to-one matching constraint. Let $\mathfrak{M}_{\max}(G_B)$ denote the set of all maximum-cardinality matchings on $G_B$. NIRA formulates subject creation relation recovery as a maximum-cardinality minimum-cost matching problem, selecting the solution with the minimum total cost among all maximum-cardinality matchings:
\begin{equation}
\hat{\mathcal{M}} = \operatornamewithlimits{arg\,min}_{\mathcal{M}' \in \mathfrak{M}_{\max}(G_B)} \sum_{(e_p, e_c) \in \mathcal{M}'} w(e_p, e_c) 
\end{equation}

Accordingly, $\hat{\mathcal{M}}$ denotes the final set of recovered subject creation relations, where $(e_p, e_c) \in \hat{\mathcal{M}}$ indicates that the parent-side and child-side records are associated with the same subject creation operation.

\subsubsection{File Descriptor Transfer Relation Recovery}
\label{subsubsec:nira_fd_transfer}
Additionally, FD transfer relationships between subjects over Unix domain sockets provide an important basis for subsequent object association. NIRA focuses on recovering FD transfers performed through \texttt{SCM\_RIGHTS} control messages, which allow the receiving subject to obtain a new FD referring to the same system resource. For observed \texttt{sendmsg} and \texttt{recvmsg} events, sender-side and receiver-side records are associated using Unix domain socket endpoints, \texttt{SCM\_RIGHTS} control messages, and temporal proximity, with candidate events satisfying $|\tau_r^{\mathrm{ret}} - \tau_s^{\mathrm{ret}}| \leq \Delta\tau_{\mathrm{ipc}}$.
When ordinary payload information is observable, its size and content are used only as auxiliary matching information rather than as strict equality constraints. For successfully associated events, an inter-subject FD transfer relationship $id_s \xrightarrow{\texttt{SCM\_RIGHTS}} id_r$ is established. If the sender-side descriptor $fd_s$ and receiver-side descriptor $fd_r$ can be further resolved, the relation $(id_s, fd_s) \xrightarrow{\mathrm{transfer}} (id_r, fd_r)$ is recorded and passed to FDProp for subsequent object association.

\subsection{FDProp}
To reconstruct syscall sequences, FDs across subjects that refer to the same underlying system resource must be consistently associated. 
Although the mapping between FD and resources changes dynamically through operations such as duplication, these operations themselves expose observable relationships among different FD references and thus provide clues for recovering the resources to which they refer. Based on this observation, we propose FDProp, which performs object association by tracking both the FD lifecycle and inter-subject descriptor propagation.
We denote the mapping between a file descriptor $fd$ in subject $u$ and its underlying system resource $o$ as $\Gamma(u,fd)=o$. FDProp continuously updates $\Gamma$ according to observable FD state changes in syscalls and FD propagation across subjects. Based on FD semantics, we categorize common operations throughout the FD lifecycle into six types: Creation, Duplication, Inheritance, Transfer, Release, and State Update, as summarized in the Appendix. For example, suppose that $fd_s$ in $u$ satisfies $\Gamma(u,fd_s)=o$. When $fd_d$ is created through a duplication operation such as \texttt{dup}, the referenced resource remains unchanged, and therefore $\Gamma(u,fd_d)=\Gamma(u,fd_s)=o$.
Inheritance typically occurs during task creation, where the newly created task either shares or copies the parent task's FD information. In this case, FDs with the same value in the parent and child tasks initially refer to the same resource, i.e., $\Gamma(u_c,fd)=\Gamma(u_p,fd)$. The subsequent effect of FD operations on $\Gamma$ depends on the task creation semantics. For example, under copy semantics, closing $fd_s$ in the child does not affect the parent task's use of the resource referenced by $fd_s$. If $fd_s$ in the child task is later reassigned to a new system resource $o'$, then $\Gamma(u_p,fd_s)=o$, $\Gamma(u_c,fd_s)=o'$, $o'\neq o$. 
In this way, FDs distributed across different subjects but referring to the same resource can be associated with a unified object identity.

\subsection{Object-centered Behavior Sequence Reconstruction}
Based on the recovered information, we dynamically reorganize the raw sequence according to the unified object identities. For any object $o_k\in\mathcal{B}$, all syscall operations satisfying $\Gamma(u_{i_j},fd_{i_j})=o_k$ under the current mapping are extracted from $\mathcal{O}$ while preserving their relative execution order, yielding
$\mathcal{O}_{o_k}=\langle op_{i_1},op_{i_2},\dots,op_{i_m}\rangle$.
The reconstructed sequence set is therefore
 $\mathcal{S}_{\mathrm{obj}}=\{\mathcal{O}_{o_1},\mathcal{O}_{o_2},\dots,\mathcal{O}_{o_K}\}$,
where each $\mathcal{O}_{o_k}$ captures the ordered operations performed on the same object and serves as the input for subsequent object-level feature extraction.

\subsection{HBSL}
To accurately capture the key semantics of malicious behavior from the reconstructed sequences, we propose HBSL, which derives subject-level behavioral representations from object-level operation features and leverages GATv2 to adaptively model relational dependencies among subjects, enabling effective learning of key behavioral patterns for HIDS.

\subsubsection{Object-level Operation Feature Extraction}
Inspired by the demonstrated effectiveness of $n$-grams in capturing local ordering and combinational patterns in prior studies~\cite{khreich2017anomaly,subba2021tfidfvectorizer,he2023br}, HBSL extracts $n$-gram features from the reconstructed sequences. The feature space $\mathcal{F}^{(n)}$ is constructed from the training set, and the sequence $\mathcal{O}_{o}$ of object $o$ is represented as $\mathbf{x}_{o} \in \mathbb{R}^{|\mathcal{F}^{(n)}|}$. To obtain a unified representation for hierarchical learning, $\mathbf{x}_{o}$ is linearly projected into a $d$-dimensional latent space as
$X_{\mathrm{seq}} = W_{\mathrm{proj}}\mathbf{x}_{o} + b_{\mathrm{proj}}$, where $X_{\mathrm{seq}} \in \mathbb{R}^{d}$,
$d=128$ in this work. This projection operates only on the feature representation and does not alter the $n$-gram feature generation process described above. To further distinguish the operational semantics associated with different object types, we introduce an object-type embedding $X_{\mathrm{meta}} \in \mathbb{R}^{d}$ and fuse it with the operation feature:
\begin{equation}
V_{\mathrm{obj}} = \operatorname{LayerNorm}\left( X_{\mathrm{seq}} + X_{\mathrm{meta}} \right)
\end{equation}

The resulting $V_{\mathrm{obj}} \in \mathbb{R}^{d}$ is used as the input to the subsequent subject-level behavioral representation stage.

\subsubsection{Subject-level Behavioral Representation}
In a complex attack, a single subject $u$ may interact with multiple objects, with operation features distributed across these objects but collectively characterizing the behavioral semantics of $u$, making effective aggregation necessary. However, our study shows that features closely associated with anomalous behavior are often concentrated on a subset of critical objects, as illustrated in Fig.~\ref{fig:top3_categories}. Treating all features equally during aggregation may weaken the model's ability to identify key behavioral patterns. Therefore, we introduce a position-encoding-based attention aggregation mechanism that adaptively weights object-level behavioral features according to their contributions, thereby producing a subject-level behavioral representation.
Specifically, for $u$, the $N$ object-level operation features form a feature matrix $H_{\mathrm{obj}} \in \mathbb{R}^{N \times d}$. To preserve the relative order of different object behaviors during subject execution, we introduce positional encoding $P \in \mathbb{R}^{N \times d}$. A two-layer MLP is then used to compute the importance score of the $i$-th object-level behavioral representation:
\begin{equation}
e_i = \mathbf{w}_2^T \tanh\left( \mathbf{W}_1 (H_{\mathrm{obj}}^{(i)} + P^{(i)}) + \mathbf{b}_1 \right) + b_2 
\end{equation}

The scores are normalized using Softmax to obtain the attention weights, $\alpha_i = \exp(e_i) / \sum_{j=1}^{N} \exp(e_j)$. The subject-level behavioral representation is then computed as:

\begin{equation}
V_{\mathrm{process}} = \sum_{i=1}^{N} \alpha_i \left( H_{\mathrm{obj}}^{(i)} + P^{(i)} \right) 
\end{equation}

This mechanism can reduce the influence of less informative object operations and emphasize behavioral patterns important for anomaly discrimination.

\subsubsection{End-to-End Detection Model}
Modern attacks often involve coordinated execution across multiple processes, with different subjects assuming different roles. Although $V_{\mathrm{process}}$ captures the behavioral characteristics of an individual subject, it cannot fully represent the dependencies among interacting subjects. Therefore, HBSL employs GATv2~\cite{brody2021attentive} to model inter-subject behavioral dependencies and adaptively aggregate information from related subjects.
Specifically, we construct a subject behavior graph $G=(V,E)$, where the initial feature of node $v_i \in V$ is $h_i^{(0)} = V_{\mathrm{process},i}$, and each edge $(v_i,v_j) \in E$ represents an inter-subject runtime relationship recovered by NIRA. Let $\mathcal{I}(i)={j \mid (v_i,v_j)\in E}$ denote the index set of neighboring nodes of $v_i$. For the $l$-th GATv2 layer, the attention score between $v_i$ and its neighbor $v_j$ is computed as:
\begin{equation}
e_{ij}^{(l)} = \mathbf{a}^{(l)T} \operatorname{LeakyReLU}\left( \mathbf{W}^{(l)} [h_i^{(l)} \Vert h_j^{(l)}] \right)
\end{equation}
The attention scores are normalized over $\mathcal{I}(i)$ using Softmax: $\alpha_{ij}^{(l)} = \exp(e_{ij}^{(l)}) / \sum_{k \in \mathcal{I}(i)} \exp(e_{ik}^{(l)})$. The neighboring representations are then aggregated as:
\begin{equation}
h_i^{(l+1)} = \sigma\left( \sum_{j \in \mathcal{I}(i)} \alpha_{ij}^{(l)} \mathbf{W}_{v}^{(l)} h_j^{(l)} \right)
\end{equation}
Two GATv2 layers are stacked to obtain relation-enhanced subject-level representations, which are then aggregated through global attention pooling. For the final-layer node representation $h_i^{(L)}$, a learnable scoring function $\psi(\cdot)$ computes the node weight $\beta_i = \exp(\psi(h_i^{(L)})) / \sum_{j=1}^{|V|} \exp(\psi(h_j^{(L)}))$. The graph-level behavioral representation is then obtained as:
\begin{equation}
Z_{\mathrm{graph}} = \sum_{i=1}^{|V|} \beta_i h_i^{(L)} 
\end{equation}

Finally, $Z_{\mathrm{graph}}$ is fed into a linear classification layer, and the prediction is obtained through Softmax: $\hat{\mathbf{y}} = \operatorname{Softmax}(\mathbf{W}_{c} Z_{\mathrm{graph}} + \mathbf{b}_{c})$. Based on the above design, our method can capture key anomalous behavioral patterns.

\section{Evaluation}
To evaluate the proposed method, we conduct experiments from four aspects: the effectiveness of sequence reconstruction, detection performance, ablation studies, and learning-curve analysis under different training-set sizes.

\subsection{Datasets and Evaluation Metrics}

Many widely used Linux HIDS datasets, such as ADFA-LD, contain only syscall names, which limits their ability to support fine-grained behavioral analysis. In addition, many datasets were collected years ago and may no longer accurately reflect current attack techniques. To address these limitations, we construct and publicly release a new Linux HIDS dataset that retains rich syscall argument information, as shown in Table~\ref{table:data_set}.
The dataset consists of benign and malicious samples. The benign portion covers 15 representative Linux application scenarios, including system utilities, file management, and other common applications.
The malicious portion includes samples spanning multiple stages of the attack lifecycle, covering attack behaviors such as privilege escalation, remote code execution, and command-and-control communication, thereby providing broad coverage of multiple tactics and techniques in the ATT\&CK framework.
We evaluate the proposed method using eight evaluation metrics: accuracy, precision, recall, F1-score, FNR, FPR, ROC-AUC, and PR-AUC.

\begin{table}[htbp]
\centering
\caption{Overview of the Dataset}

\renewcommand{\arraystretch}{1.2}

\begin{tabularx}{\linewidth}{l >{\raggedright\arraybackslash}X l}
\hline
\textbf{Component} & \textbf{Description} & \textbf{Specification} \\
\hline

Collection Tool 
& Runtime syscall monitoring 
& Sysdig v0.40.1 \\

Data Type 
& Syscall traces with arguments 
& Runtime-level \\

Benign Samples 
& Daily Linux user activities 
& 1309 samples \\

Malicious Samples 
& CWE-, CVE-, and malware-based attacks 
& 329 samples \\

Attack Coverage 
& TTP-oriented multi-stage attacks 
& ATT\&CK-based \\

\hline
\label{table:data_set}
\end{tabularx}
\end{table}

\subsection{Environment}
The method is implemented using PyTorch and PyTorch Geometric. All experiments are conducted on Ubuntu 22.04.5 LTS with an Intel(R) Core(TM) i9-12900KF CPU and an NVIDIA GeForce RTX 4090 GPU. To ensure fairness and reproducibility, the model's major hyperparameters follow prior security studies and commonly used deep learning configurations, without extensive dataset-specific tuning.

\subsection{Analysis of Reconstructed Behavior Sequences}
\subsubsection{Behavior Sequence Reconstruction}
\begin{figure}[htbp]
    \centering
    \includegraphics[width=1.0\columnwidth]{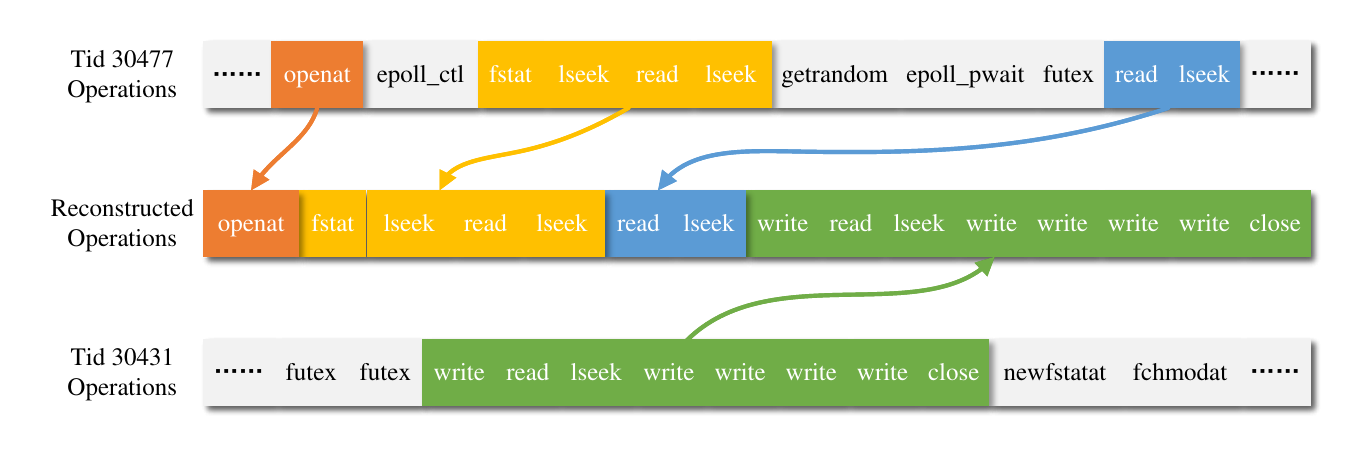} 
    \caption{Example of Behavior Sequence Reconstruction.}
    \label{fig:example_pic}
\end{figure}

Taking a publicly available ransomware sample from GitHub\footnote{https://github.com/vxunderground/MalwareSourceCode.git} as an example, \texttt{fd=3} (\path{/home/glibc-2.39/benchtests/Makefile}) is accessed successively by threads 30477 and 30431 within the same thread group 30400. Thread 30477 performs read-related operations such as \texttt{openat}, \texttt{fstat}, \texttt{lseek}, and \texttt{read}, while thread 30431 performs multiple write-related operations, including \texttt{write} and \texttt{lseek}. In the raw sequences, these related operations are interleaved with concurrent activities involving other subjects and objects, fragmenting their behavioral semantics. As shown in Fig.~\ref{fig:example_pic}, the reconstructed sequence reconnects the related operations distributed across different tasks, making the complete operation process clearly observable, from retrieving the original file contents to repeatedly overwriting and updating the file. Compared with the interleaved pattern in the raw sequence, the reconstructed sequence presents the underlying malicious behavioral semantics more clearly and intuitively.

\subsubsection{$n$-gram Feature Extraction}
Compared with feature extraction from the raw sequences and the MGFE~\cite{he2023br} method, feature extraction based on the reconstructed syscall sequences substantially reduces the number of features introduced by semantic fragmentation, while preserving more continuous behavioral semantics and critical information in the extracted $n$-grams. Specifically, for $n$-gram lengths of 2, 3, 4, 5, 6, and 7, our method reduces the number of features by 2,368, 7,320, 28,330, 66,842, 139,582, and 237,094, corresponding to reductions of 48.22\%, 49.07\%, 62.51\%, 70.32\%, 76.56\%, and 79.78\%, respectively, compared with the raw-sequence method. Compared with MGFE, our method reduces the total number of extracted features by 125,207, corresponding to an overall reduction of 44.1\%.

\begin{figure}[htbp]
    \centering
    \includegraphics[width=0.9\columnwidth]{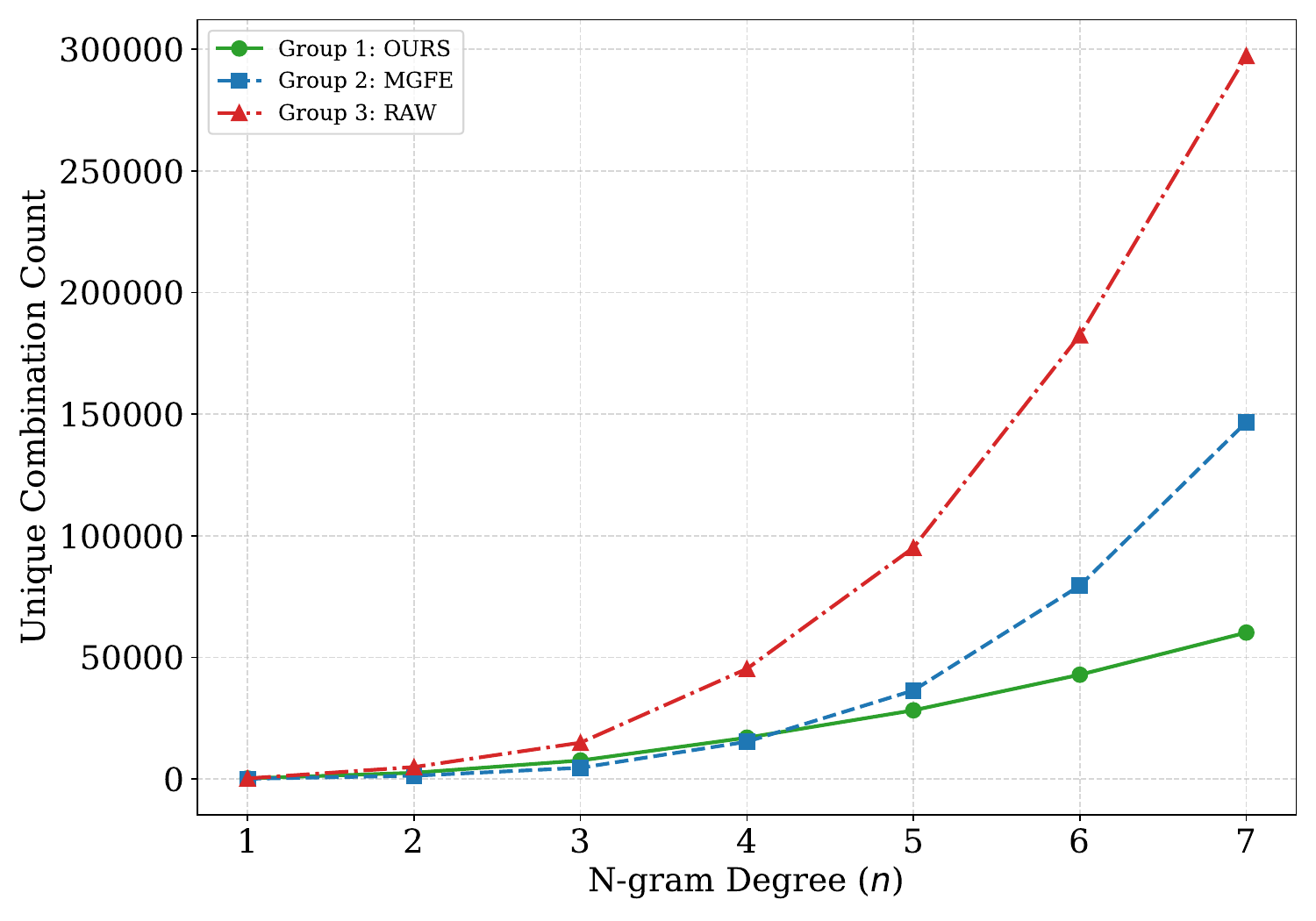} 
    \caption{Comparison of $n$-gram feature counts across different methods.}
    \label{fig:ngram_single_column}
\end{figure}

\subsubsection{Detection Performance of $n$-gram Features }

To evaluate the behavior sequences reconstructed by the proposed method, we compare the detection performance of $n$-gram features extracted before and after sequence reconstruction. Table~\ref{tab:performance_time_comparison} presents the comparative results under different intrusion detection metrics. Using four representative machine learning classifiers, namely SVM, XGBoost, KNN, and DT, the reconstructed sequences (OURS) are compared with the raw sequences across 28 metric combinations (7 term sizes $\times$ 4 evaluation metrics). OURS achieves better results in 17, 24, 26, and 19 of these combinations, respectively.
The results show that, as the term size increases, the detection performance of the raw-sequence-based method declines noticeably due to the increasing number of irrelevant features. This effect is particularly evident for KNN, which performs classification based on distances in the feature space and is therefore more sensitive to the stability and similarity of feature distributions. In contrast, the reconstructed behavior sequences improve behavioral semantic continuity and the density of relevant information while substantially reducing the number of irrelevant features, resulting in clear overall advantages across multiple metrics, including Precision, Recall, F1-score, and FPR, under the same classifiers. For example, under the 4-gram setting, the F1-score of KNN increases from 0.2949 for the Baseline to 0.7692 for OURS. In addition, the proposed method achieves lower detection time across different classifiers.

\begin{table*}[t]
\centering
\caption{Comparison of detection performance and training/testing time across different classifiers and $n$-gram sizes}
\label{tab:performance_time_comparison}

\setlength{\tabcolsep}{3pt}
\renewcommand{\arraystretch}{0.65}

\resizebox{0.9\textwidth}{!}{%
\begin{tabular}{lccccccccccccc}
\toprule
\multirow{2}{*}{\textbf{Classifiers}}
& \multirow{2}{*}{\textbf{Term-size}}
& \multicolumn{6}{c}{\textbf{Raw N-gram}}
& \multicolumn{6}{c}{\textbf{OUR N-gram}} \\
\cmidrule(lr){3-8}
\cmidrule(lr){9-14}

& 
& Training time (s)
& Testing time (s)
& Precision
& Recall
& F1-score
& FPR
& Training time (s)
& Testing time (s)
& Precision
& Recall
& F1-score
& FPR \\
\midrule

\multirow{7}{*}{SVM}
& 1
& 0.2747 & 0.0211 & 0.8842 & 0.9032 & 0.8936 & 0.0280
& \textbf{0.1311} $\downarrow$
& \textbf{0.0101} $\downarrow$
& \textbf{0.9079} $\uparrow$
& 0.7500
& 0.8214
& \textbf{0.0178} $\downarrow$ \\

& 2
& 0.4583 & 0.0335 & 0.8767 & 0.6882 & 0.7711 & 0.0229
& \textbf{0.1542} $\downarrow$
& \textbf{0.0128} $\downarrow$
& 0.7901
& \textbf{0.6957} $\uparrow$
& 0.7399
& 0.0433 \\

& 3
& 2.6962 & 0.1721 & 0.8644 & 0.5484 & 0.6711 & 0.0204
& \textbf{0.4880} $\downarrow$
& \textbf{0.0371} $\downarrow$
& 0.8256
& \textbf{0.7717} $\uparrow$
& \textbf{0.7978} $\uparrow$
& 0.0382 \\

& 4
& 3.9414 & 0.2546 & 0.4463 & 0.8495 & 0.5852 & 0.2494
& \textbf{0.3739} $\downarrow$
& \textbf{0.0366} $\downarrow$
& \textbf{0.5725} $\uparrow$
& 0.8152
& \textbf{0.6726} $\uparrow$
& \textbf{0.1425} $\downarrow$ \\

& 5
& 6.2752 & 0.4397 & 0.2529 & 0.9355 & 0.3982 & 0.6540
& \textbf{0.2838} $\downarrow$
& \textbf{0.0385} $\downarrow$
& \textbf{0.2757} $\uparrow$
& 0.9022
& \textbf{0.4224} $\uparrow$
& \textbf{0.5549} $\downarrow$ \\

& 6
& 7.9534 & 0.6564 & 0.2557 & 0.9570 & 0.4036 & 0.6592
& \textbf{0.2825} $\downarrow$
& \textbf{0.0431} $\downarrow$
& \textbf{0.2926} $\uparrow$
& 0.8587
& \textbf{0.4365} $\uparrow$
& \textbf{0.4860} $\downarrow$ \\

& 7
& 12.2457 & 1.1130 & 0.2602 & 0.9570 & 0.4092 & 0.6439
& \textbf{0.3210} $\downarrow$
& \textbf{0.0479} $\downarrow$
& \textbf{0.2706} $\uparrow$
& 0.8913
& \textbf{0.4152} $\uparrow$
& \textbf{0.5624} $\downarrow$ \\

\midrule

\multirow{7}{*}{XGBoost}
& 1
& 0.9257 & 0.0025 & 0.9247 & 0.9247 & 0.9247 & 0.0178
& 1.1978
& \textbf{0.0024} $\downarrow$
& \textbf{0.9390} $\uparrow$
& 0.8370
& 0.8851
& \textbf{0.0127} $\downarrow$ \\

& 2
& 0.1892 & 0.0052 & 0.9054 & 0.7204 & 0.8024 & 0.0178
& \textbf{0.1334} $\downarrow$
& \textbf{0.0030} $\downarrow$
& \textbf{0.9103} $\uparrow$
& \textbf{0.7717} $\uparrow$
& \textbf{0.8353} $\uparrow$
& 0.0178 \\

& 3
& 2.7246 & 0.0346 & 0.9359 & 0.7849 & 0.8538 & 0.0127
& \textbf{0.1422} $\downarrow$
& \textbf{0.0050} $\downarrow$
& \textbf{1.0000} $\uparrow$
& \textbf{0.8587} $\uparrow$
& \textbf{0.9240} $\uparrow$
& \textbf{0.0000} $\downarrow$ \\

& 4
& 5.1258 & 0.1056 & 0.8916 & 0.7957 & 0.8409 & 0.0229
& \textbf{0.2933} $\downarrow$
& \textbf{0.0097} $\downarrow$
& \textbf{0.9136} $\uparrow$
& \textbf{0.8043} $\uparrow$
& \textbf{0.8555} $\uparrow$
& \textbf{0.0178} $\downarrow$ \\

& 5
& 10.8975 & 0.3964 & 0.8732 & 0.6667 & 0.7561 & 0.0229
& \textbf{0.4190} $\downarrow$
& \textbf{0.0113} $\downarrow$
& \textbf{0.9022} $\uparrow$
& \textbf{0.9022} $\uparrow$
& \textbf{0.9022} $\uparrow$
& 0.0229 \\

& 6
& 20.3514 & 0.9731 & 0.9649 & 0.5914 & 0.7333 & 0.0051
& \textbf{0.5474} $\downarrow$
& \textbf{0.0139} $\downarrow$
& \textbf{1.0000} $\uparrow$
& \textbf{0.8261} $\uparrow$
& \textbf{0.9048} $\uparrow$
& \textbf{0.0000} $\downarrow$ \\

& 7
& 33.8241 & 1.8814 & 0.7805 & 0.6882 & 0.7314 & 0.0458
& \textbf{1.7005} $\downarrow$
& \textbf{0.0216} $\downarrow$
& \textbf{0.9733} $\uparrow$
& \textbf{0.7935} $\uparrow$
& \textbf{0.8743} $\uparrow$
& \textbf{0.0051} $\downarrow$ \\

\midrule

\multirow{7}{*}{KNN}
& 1
& 0.0014 & 0.0815 & 0.9043 & 0.9140 & 0.9091 & 0.0229
& 0.0015
& \textbf{0.0615} $\downarrow$
& \textbf{0.9157} $\uparrow$
& 0.8261
& 0.8686
& \textbf{0.0178} $\downarrow$ \\

& 2
& 0.0016 & 0.0726 & 0.7838 & 0.6237 & 0.6946 & 0.0407
& \textbf{0.0013} $\downarrow$
& \textbf{0.0625} $\downarrow$
& \textbf{0.8488} $\uparrow$
& \textbf{0.7935} $\uparrow$
& \textbf{0.8202} $\uparrow$
& \textbf{0.0331} $\downarrow$ \\

& 3
& 0.0022 & 0.1340 & 0.4933 & 0.3978 & 0.4405 & 0.0967
& \textbf{0.0015} $\downarrow$
& \textbf{0.0672} $\downarrow$
& \textbf{0.9467} $\uparrow$
& \textbf{0.7717} $\uparrow$
& \textbf{0.8503} $\uparrow$
& \textbf{0.0102} $\downarrow$ \\

& 4
& 0.0027 & 0.2194 & 0.2581 & 0.3441 & 0.2949 & 0.2341
& \textbf{0.0015} $\downarrow$
& \textbf{0.0687} $\downarrow$
& \textbf{0.7778} $\uparrow$
& \textbf{0.7609} $\uparrow$
& \textbf{0.7692} $\uparrow$
& \textbf{0.0509} $\downarrow$ \\

& 5
& 0.0039 & 0.4453 & 0.3594 & 0.2473 & 0.2930 & 0.1043
& \textbf{0.0016} $\downarrow$
& \textbf{0.0801} $\downarrow$
& \textbf{0.5625} $\uparrow$
& \textbf{0.4891} $\uparrow$
& \textbf{0.5233} $\uparrow$
& \textbf{0.0891} $\downarrow$ \\

& 6
& 0.0220 & 0.8522 & 0.3393 & 0.2043 & 0.2550 & 0.0941
& \textbf{0.0017} $\downarrow$
& \textbf{0.0814} $\downarrow$
& \textbf{0.7353} $\uparrow$
& \textbf{0.2717} $\uparrow$
& \textbf{0.3968} $\uparrow$
& \textbf{0.0229} $\downarrow$ \\

& 7
& 0.0122 & 1.6344 & 0.2394 & 0.1828 & 0.2073 & 0.1374
& \textbf{0.0017} $\downarrow$
& \textbf{0.0846} $\downarrow$
& \textbf{0.6970} $\uparrow$
& \textbf{0.2500} $\uparrow$
& \textbf{0.3680} $\uparrow$
& \textbf{0.0254} $\downarrow$ \\

\midrule

\multirow{7}{*}{DT}
& 1
& 0.0075 & 0.0005 & 0.8804 & 0.8710 & 0.8757 & 0.0280
& \textbf{0.0043} $\downarrow$
& \textbf{0.0004} $\downarrow$
& \textbf{0.8953} $\uparrow$
& 0.8370
& 0.8652
& \textbf{0.0229} $\downarrow$ \\

& 2
& 0.0567 & 0.0005 & 0.8333 & 0.7527 & 0.7910 & 0.0356
& \textbf{0.0106} $\downarrow$
& \textbf{0.0004} $\downarrow$
& \textbf{0.8605} $\uparrow$
& \textbf{0.8043} $\uparrow$
& \textbf{0.8315} $\uparrow$
& \textbf{0.0305} $\downarrow$ \\

& 3
& 0.4979 & 0.0015 & 0.9070 & 0.8387 & 0.8715 & 0.0204
& \textbf{0.0468} $\downarrow$
& \textbf{0.0006} $\downarrow$
& \textbf{1.0000} $\uparrow$
& 0.7826
& \textbf{0.8780} $\uparrow$
& \textbf{0.0000} $\downarrow$ \\

& 4
& 1.7177 & 0.0035 & 0.7802 & 0.7634 & 0.7717 & 0.0509
& \textbf{0.0782} $\downarrow$
& \textbf{0.0008} $\downarrow$
& \textbf{0.7979} $\uparrow$
& \textbf{0.8152} $\uparrow$
& \textbf{0.8065} $\uparrow$
& \textbf{0.0483} $\downarrow$ \\

& 5
& 6.0927 & 0.0112 & 0.8608 & 0.7312 & 0.7907 & 0.0280
& \textbf{0.1434} $\downarrow$
& \textbf{0.0009} $\downarrow$
& \textbf{0.9344} $\uparrow$
& 0.6196
& 0.7451
& \textbf{0.0102} $\downarrow$ \\

& 6
& 17.9611 & 0.0209 & 0.8906 & 0.6129 & 0.7261 & 0.0178
& \textbf{0.2796} $\downarrow$
& \textbf{0.0011} $\downarrow$
& \textbf{0.9000} $\uparrow$
& 0.3913
& 0.5455
& \textbf{0.0102} $\downarrow$ \\

& 7
& 26.9502 & 0.0382 & 0.7444 & 0.7204 & 0.7322 & 0.0585
& \textbf{0.4297} $\downarrow$
& \textbf{0.0013} $\downarrow$
& \textbf{1.0000} $\uparrow$
& 0.4022
& 0.5736
& \textbf{0.0000} $\downarrow$ \\

\bottomrule
\end{tabular}%
}

\end{table*}

\subsection{Object-Type Importance Across Attack Categories}
\begin{figure}[t]
    \centering
    \includegraphics[width=0.9\columnwidth]{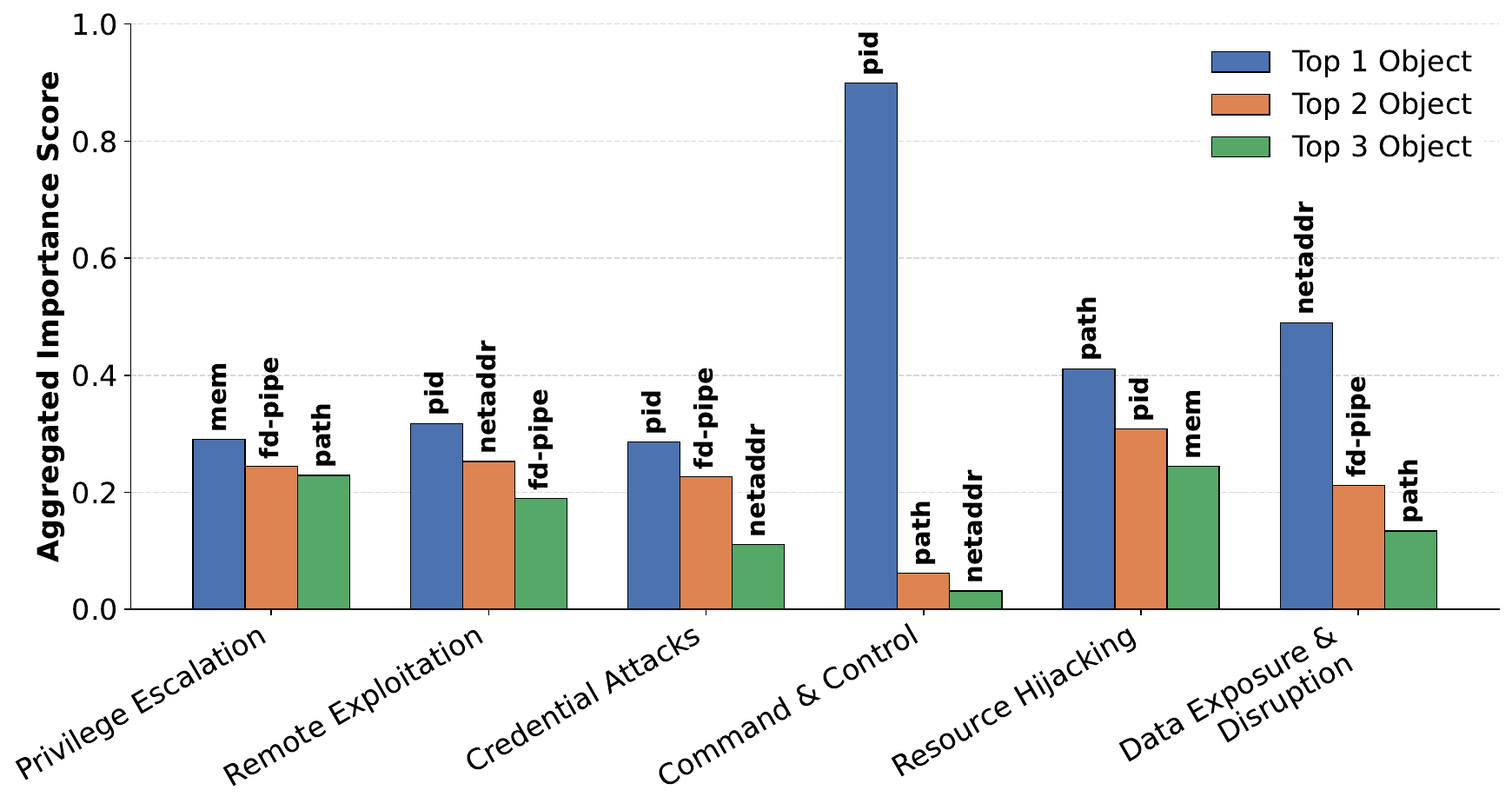}
    \caption{Top-3 important object categories across malicious behavior categories.}
    \label{fig:top3_categories}
\end{figure}

The sequence features extracted in the previous section correspond to behavioral features associated with different objects. During our study, we observed that, in the intrusion detection scenario, the importance of different object types varies across different categories of samples, as shown in Fig.~\ref{fig:top3_categories}. For example, in covert command-and-control scenarios, i.e., C2 beacons, the 'pid' object plays a dominant role, with an importance score as high as 0.9, indicating its high-frequency interaction characteristics in maintaining stealthy communication. In data exfiltration and destructive behaviors, however, the network address object 'netaddr' becomes the most important feature, with a score close to 0.5, which is highly consistent with the physical semantics of outward data transmission. In addition, local privilege escalation (LPE) tends to rely more heavily on operations related to memory objects ('mem'), whereas remote code execution (RCE) exhibits the strongest feature dependency on process objects ('pid'). This motivates the use of attention-based aggregation to adaptively capture the contributions of behaviors associated with different object categories, while GATv2 further models the collaborative behavioral patterns among interacting subjects.

\subsection{result}
\subsubsection{Comparison Results of Detection Performance}

\begin{figure*}[t]
    \centering

    \begin{minipage}{0.24\textwidth}
        \centering
        \includegraphics[width=\linewidth]{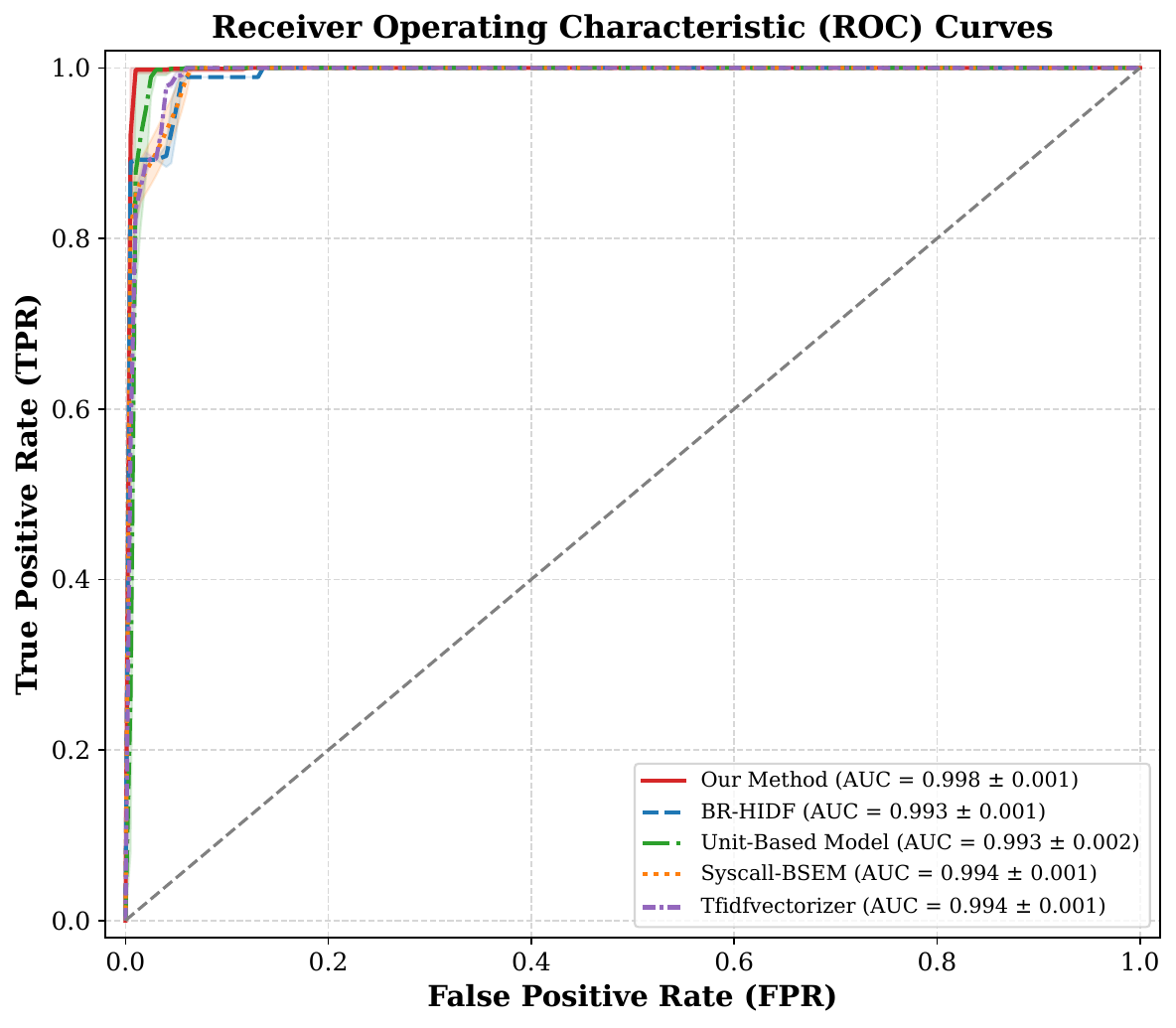}
        \centerline{\raisebox{0.4em}{\footnotesize (a)}}
    \end{minipage}%
    \hspace{0.005\textwidth}%
    \begin{minipage}{0.24\textwidth}
        \centering
        \includegraphics[width=\linewidth]{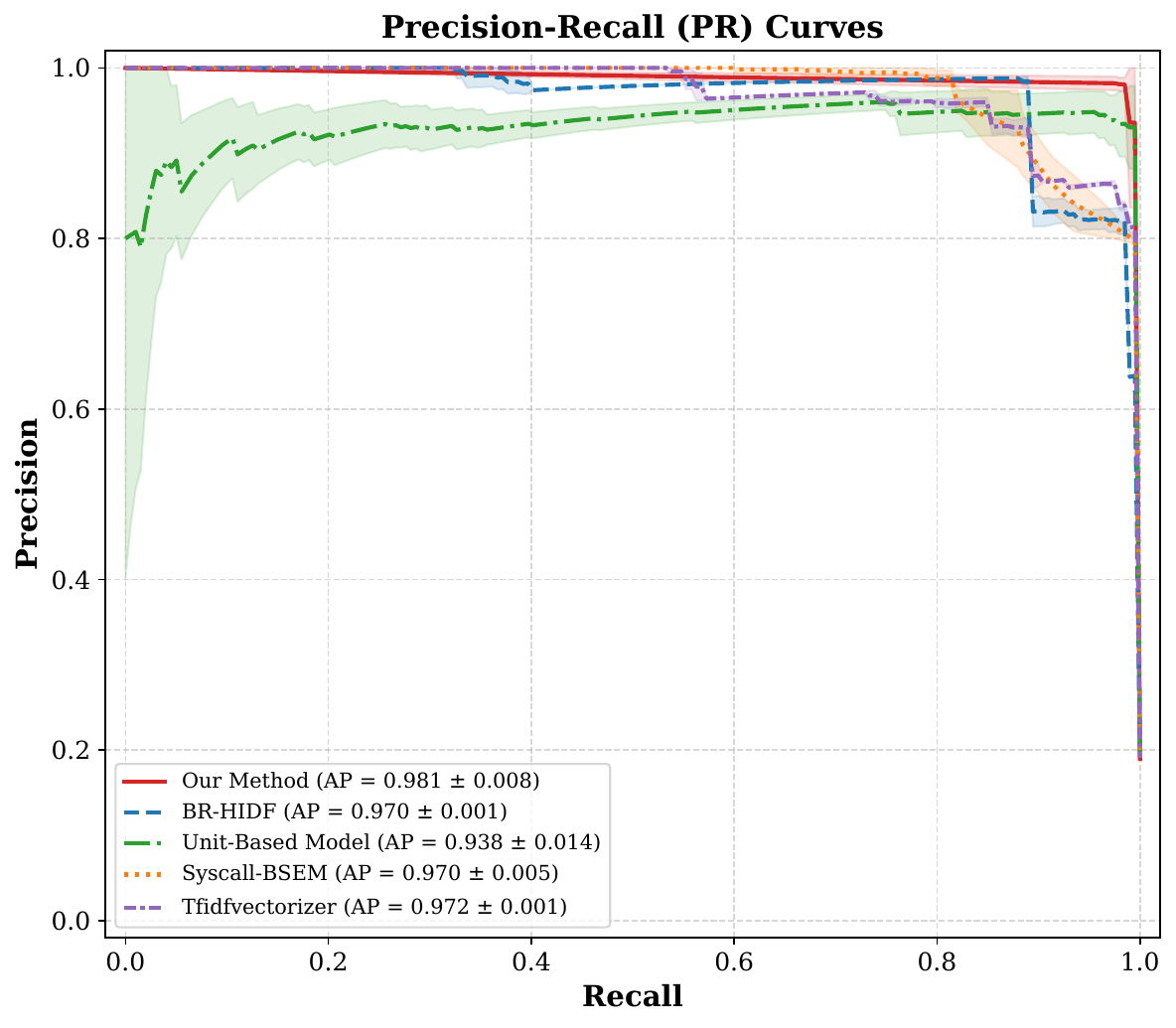}
        \centerline{\raisebox{0.4em}{\footnotesize (b)}}
    \end{minipage}%
    \hspace{0.005\textwidth}%
    \begin{minipage}{0.24\textwidth}
        \centering
        \includegraphics[width=\linewidth]{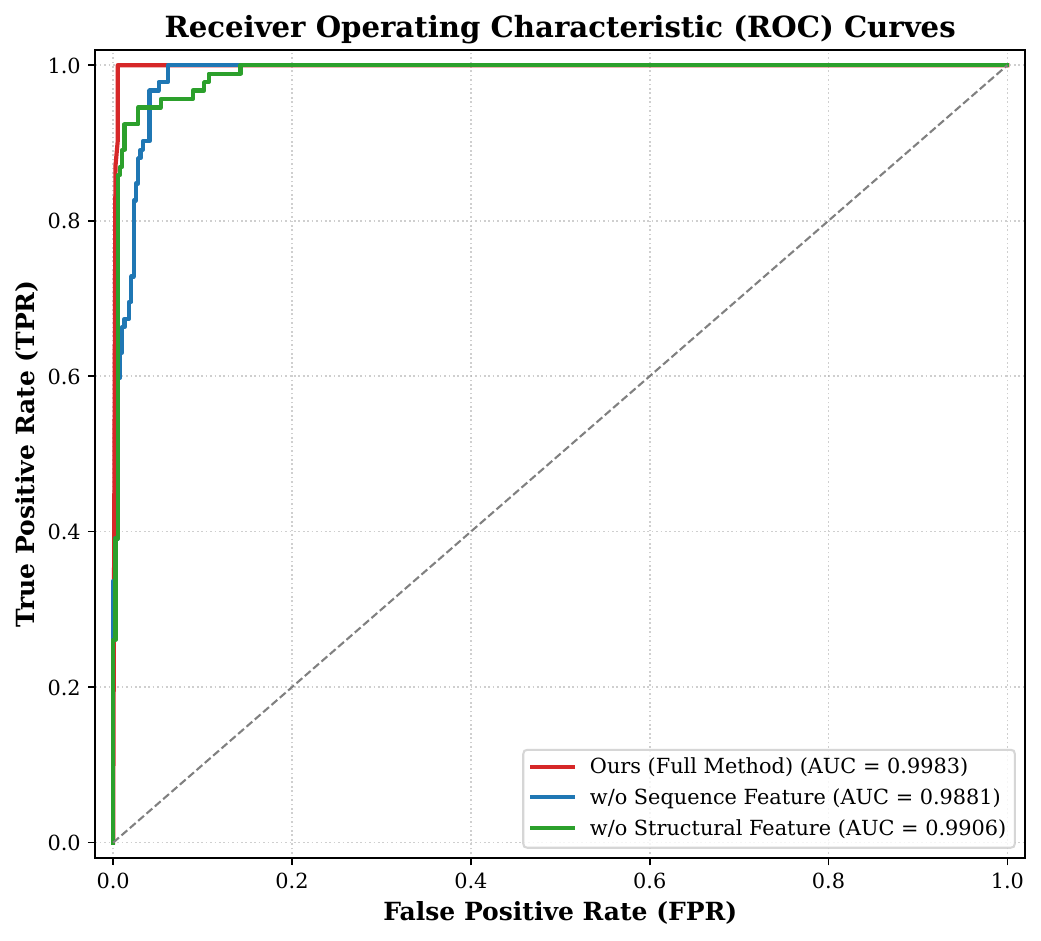}
        \centerline{\raisebox{0.4em}{\footnotesize (c)}}
    \end{minipage}%
    \hspace{0.005\textwidth}%
    \begin{minipage}{0.24\textwidth}
        \centering
        \includegraphics[width=\linewidth]{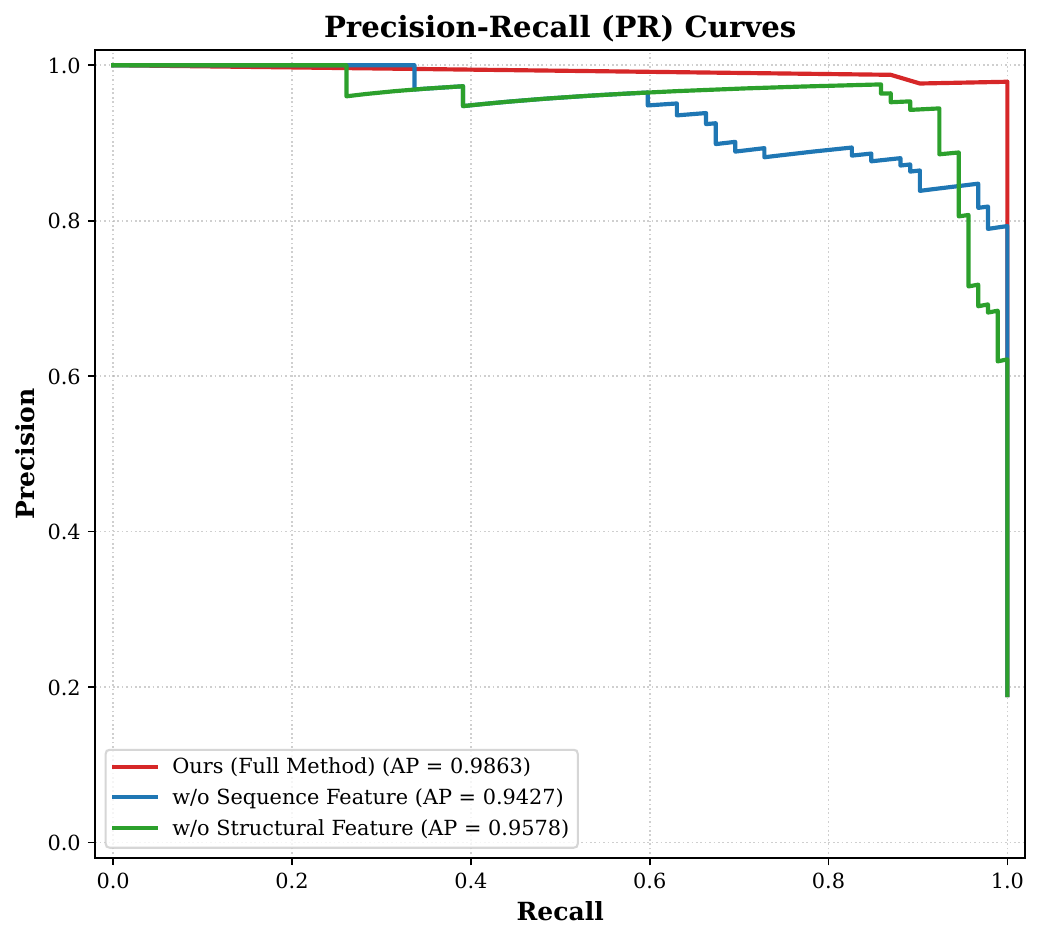}
        \centerline{\raisebox{0.4em}{\footnotesize (d)}}
    \end{minipage}

    \caption{ROC and PR curve comparisons for the HIDS models and ablation study: (a) ROC curves of the compared models, (b) PR curves of the compared models, (c) ROC curves of the ablation study, and (d) PR curves of the ablation study.}
    \label{fig:roc_pr_comparison}
\end{figure*}

Table~\ref{tab:performance_comparison} reports the average detection performance of the proposed method and four existing HIDS methods over 10 independent runs. The proposed method achieves the highest Accuracy and F1-score, reaching 0.9948 and 0.9864, respectively, while its Precision and Recall are 0.9807 and 0.9923, both ranking second among all methods. The corresponding FNR and FPR are 0.0077 and 0.0046, respectively. In addition, as shown in Fig.~\ref{fig:roc_pr_comparison} (a) and (b), our method achieves ROC-AUC and PR-AUC values of 0.998 and 0.981, respectively, both of which are the highest among all compared methods. These results indicate strong overall class separability and effective identification of malicious samples. Notably, our method uses only a linear classification layer for the final prediction, suggesting that the preceding behavior sequence reconstruction and hierarchical feature learning stages are able to produce highly discriminative behavioral representations.

In terms of individual metrics, BR-HIDF achieves the highest Precision of 0.9846, but its Recall is only 0.8924, corresponding to an FNR of 0.1076. 
This result may be related to its relatively coarse-grained behavioral representation, where stronger behavioral abstraction helps suppress false positives but may also discard some fine-grained attack information, thereby favoring Precision at the expense of Recall.
In contrast, the Unit-Based Model achieves the highest Recall of 0.9946, but its Precision decreases to 0.9447 and its FPR increases to 0.0141. This may be associated with its behavior-unit-based representation, which preserves more behavioral details and improves the robustness of detection, but may also increase sensitivity to behavior units that are insufficiently covered during training, resulting in more false positives.
In comparison, our method achieves the best Accuracy, F1-score, ROC-AUC, and PR-AUC, while ranking second in Precision, Recall, FNR, and FPR with only small gaps from the corresponding best results, demonstrating a more balanced and competitive overall detection performance.

\begin{figure*}[t]
    \centering

    \begin{minipage}{0.194\textwidth}
        \centering
        \includegraphics[width=\linewidth]{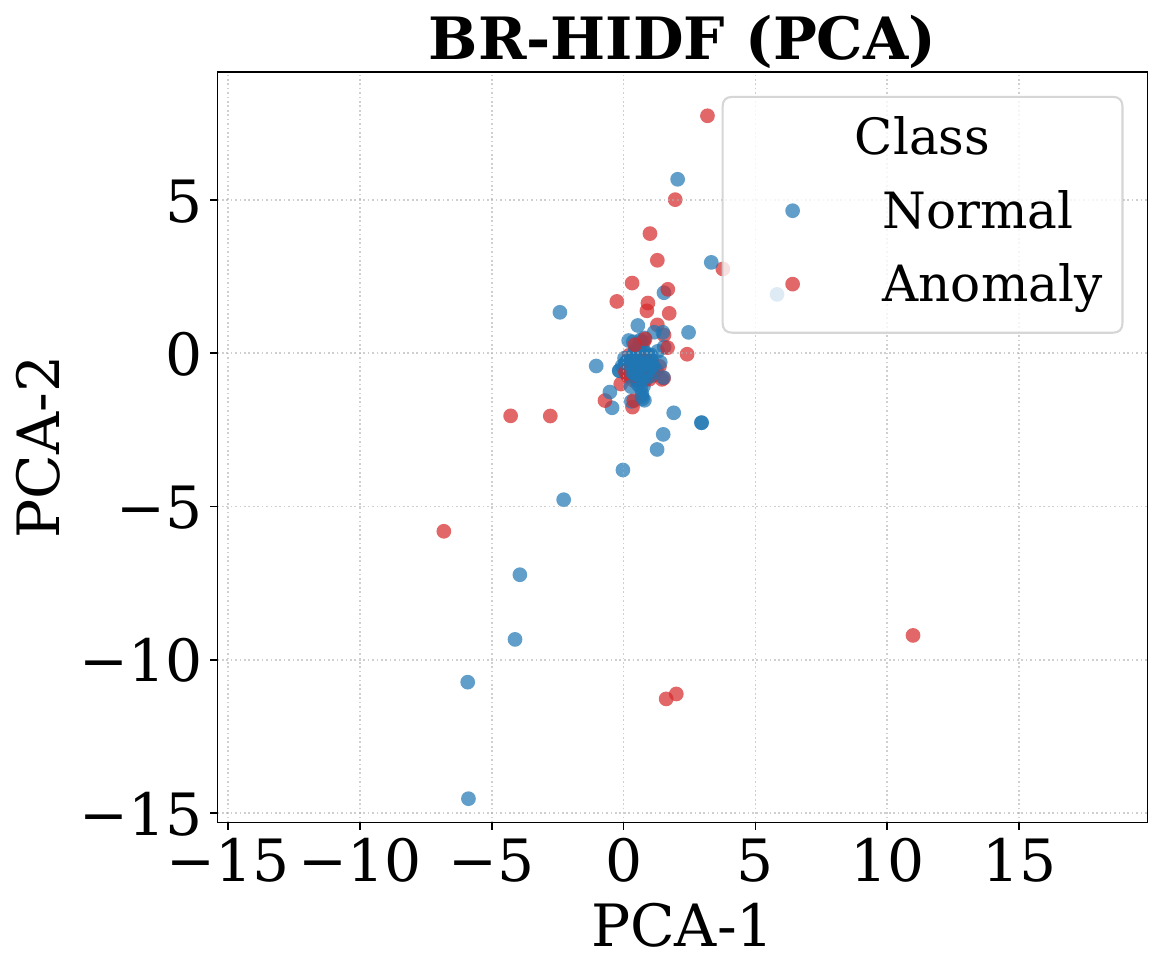}
        \centerline{\raisebox{0.4em}{\footnotesize (a)}}
    \end{minipage}%
    \hspace{0.005\textwidth}%
    \begin{minipage}{0.194\textwidth}
        \centering
        \includegraphics[width=\linewidth]{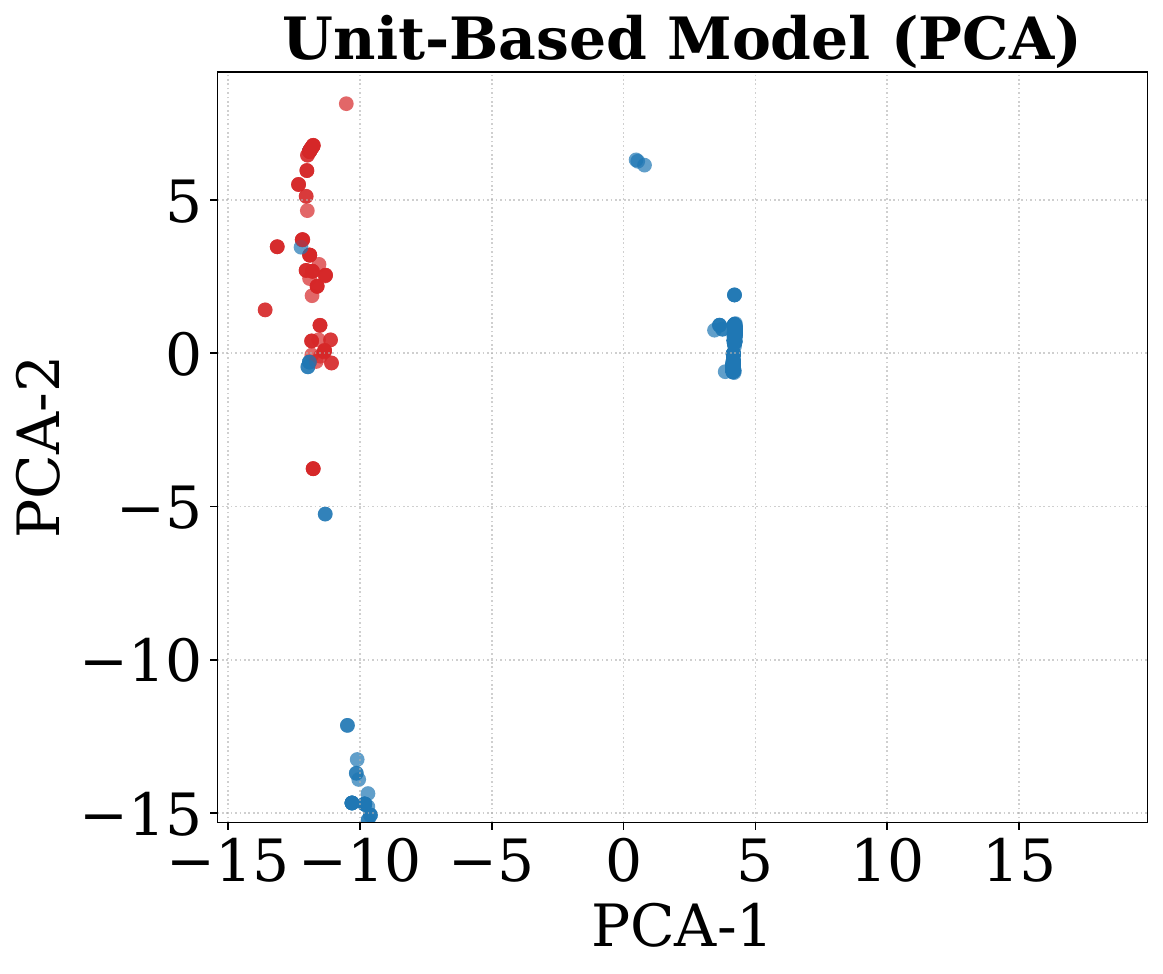}
        \centerline{\raisebox{0.4em}{\footnotesize (b)}}
    \end{minipage}%
    \hspace{0.005\textwidth}%
    \begin{minipage}{0.194\textwidth}
        \centering
        \includegraphics[width=\linewidth]{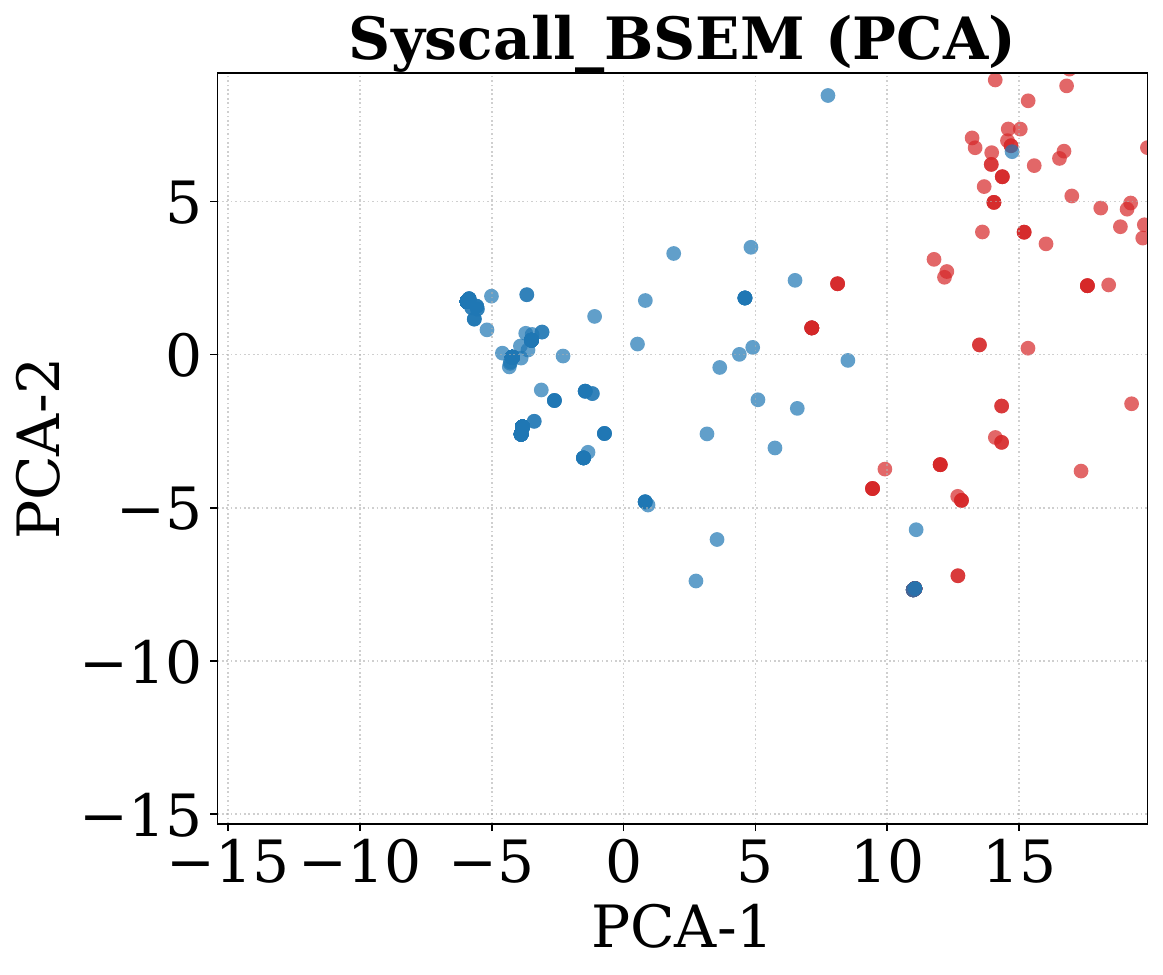}
        \centerline{\raisebox{0.4em}{\footnotesize (c)}}
    \end{minipage}%
    \hspace{0.005\textwidth}%
    \begin{minipage}{0.194\textwidth}
        \centering
        \includegraphics[width=\linewidth]{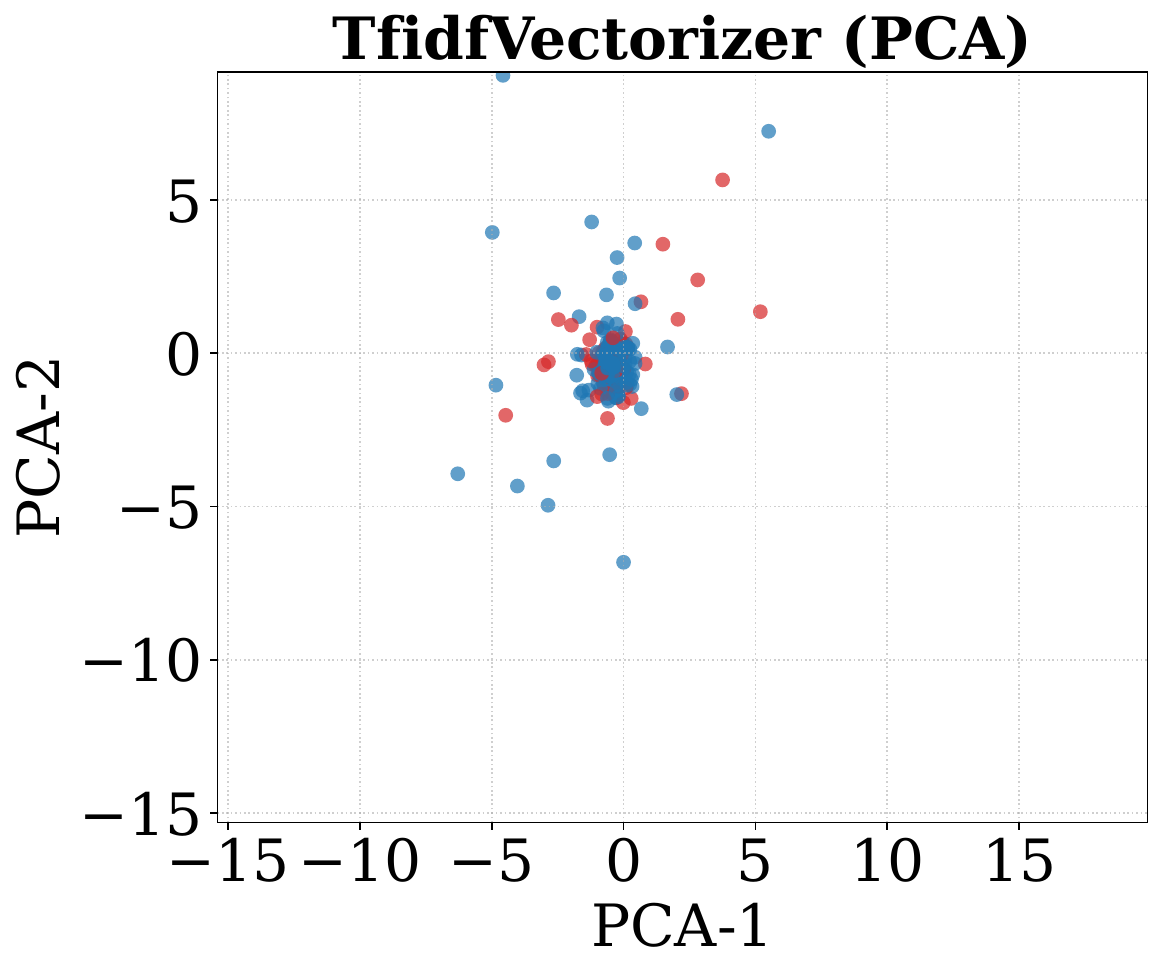}
        \centerline{\raisebox{0.4em}{\footnotesize (d)}}
    \end{minipage}%
    \hspace{0.005\textwidth}%
    \begin{minipage}{0.194\textwidth}
        \centering
        \includegraphics[width=\linewidth]{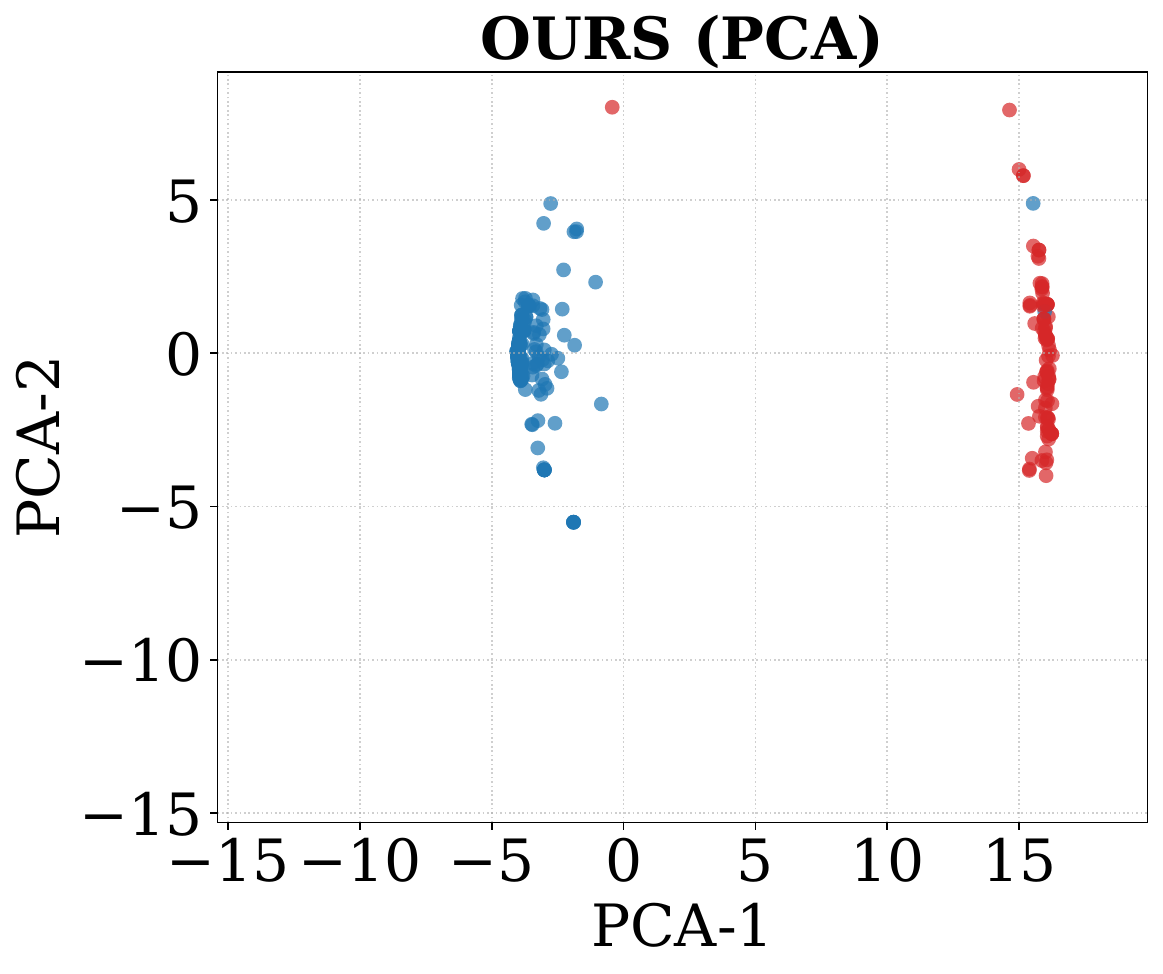}
        \centerline{\raisebox{0.4em}{\footnotesize (e)}}
    \end{minipage}

    \vspace{1em}

    \begin{minipage}{0.194\textwidth}
        \centering
        \includegraphics[width=\linewidth]{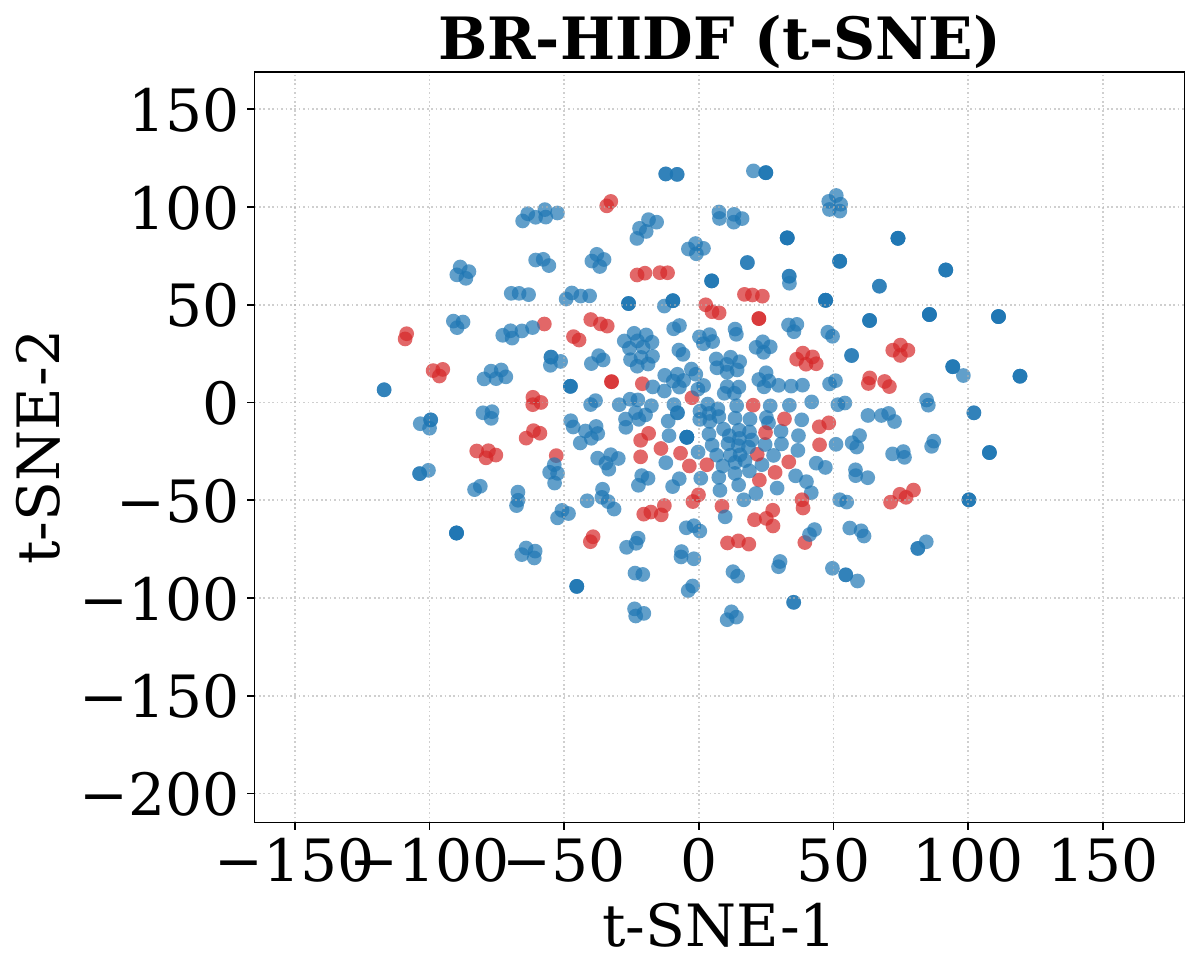}
        \centerline{\raisebox{0.4em}{\footnotesize (f)}}
    \end{minipage}%
    \hspace{0.005\textwidth}%
    \begin{minipage}{0.194\textwidth}
        \centering
        \includegraphics[width=\linewidth]{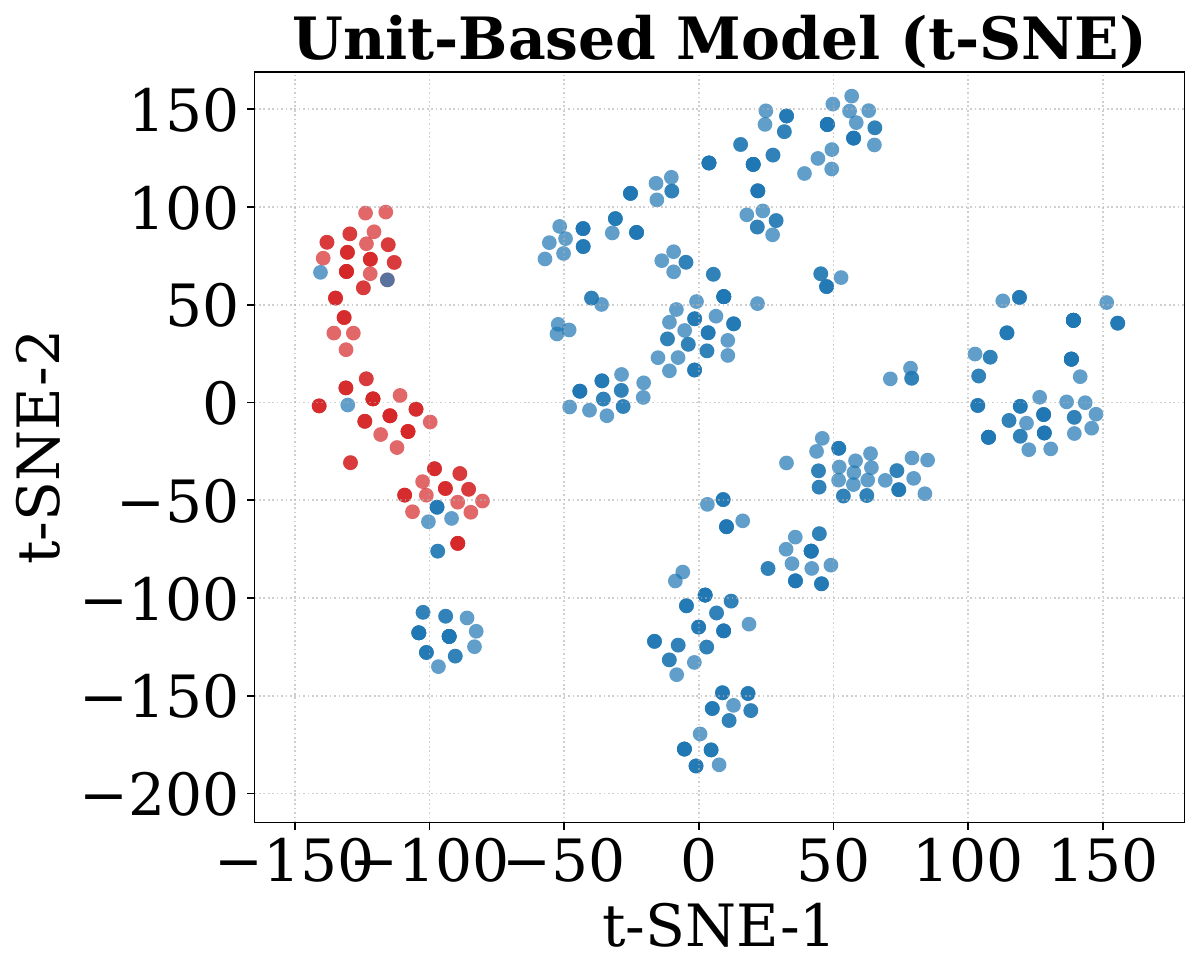}
        \centerline{\raisebox{0.4em}{\footnotesize (g)}}
    \end{minipage}%
    \hspace{0.005\textwidth}%
    \begin{minipage}{0.194\textwidth}
        \centering
        \includegraphics[width=\linewidth]{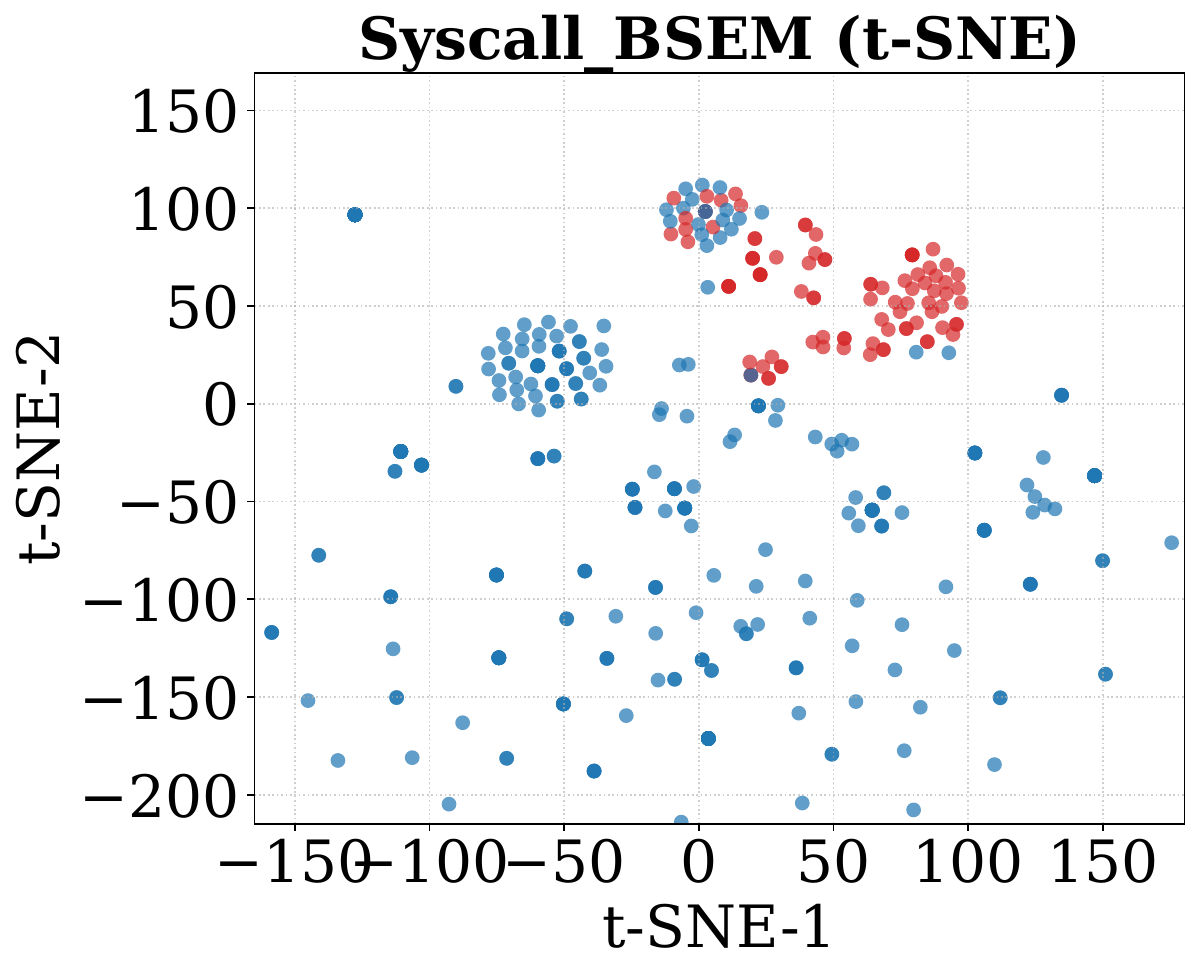}
        \centerline{\raisebox{0.4em}{\footnotesize (h)}}
    \end{minipage}%
    \hspace{0.005\textwidth}%
    \begin{minipage}{0.194\textwidth}
        \centering
        \includegraphics[width=\linewidth]{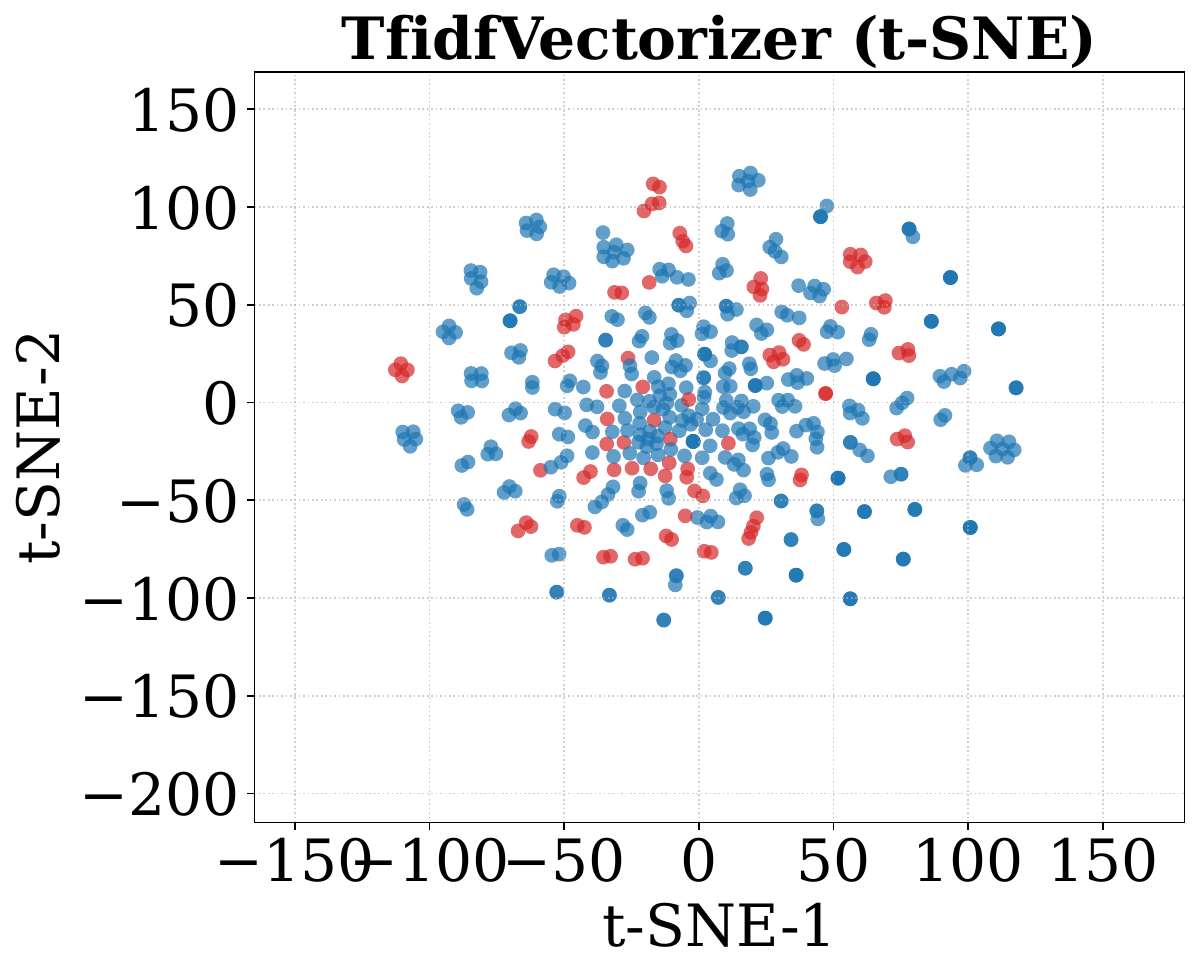}
        \centerline{\raisebox{0.4em}{\footnotesize (i)}}
    \end{minipage}%
    \hspace{0.005\textwidth}%
    \begin{minipage}{0.194\textwidth}
        \centering
        \includegraphics[width=\linewidth]{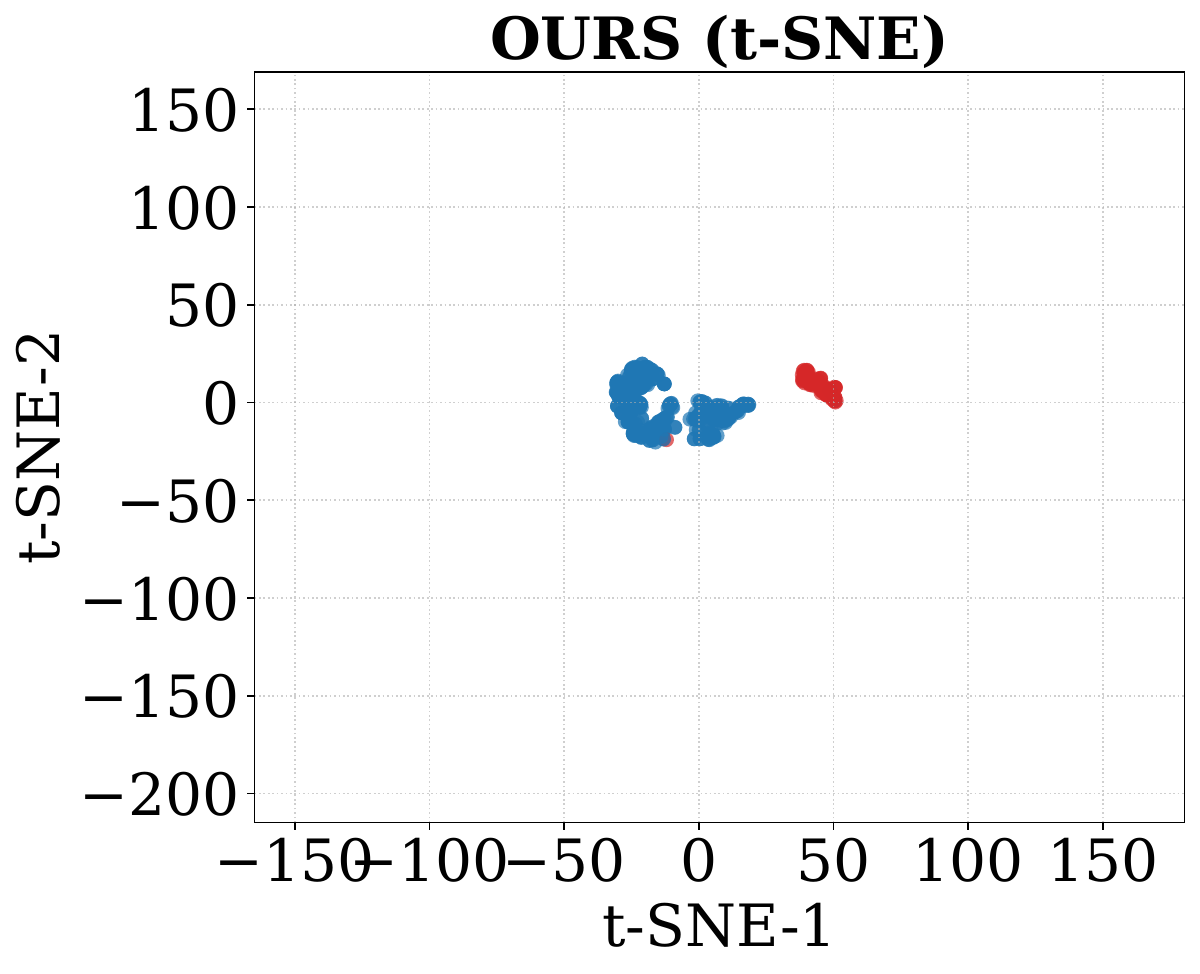}
        \centerline{\raisebox{0.4em}{\footnotesize (j)}}
    \end{minipage}

    \caption{Comparison of PCA (top row) and t-SNE (bottom row) visualizations for different models.}
    \label{fig:full_comparison}
\end{figure*}

To further examine the class distributions of the features learned by different methods, we use PCA and t-SNE to project the extracted high-dimensional features into two-dimensional space. PCA is used to examine the global linear distribution of the features, while t-SNE is used to visualize their local neighborhood structure. As shown in Fig.~\ref{fig:full_comparison}, compared with the other methods, the benign and malicious features extracted by our method exhibit clearer class separation and less overlap in both the PCA and t-SNE spaces, further indicating that the learned behavioral representations have stronger class-discriminative capability.

\begin{table}[t]
\centering
\footnotesize 
\setlength{\tabcolsep}{0pt} 
\caption{Performance Comparison of Intrusion Detection Models}
\label{tab:performance_comparison}
\begin{tabular*}{\columnwidth}{@{\extracolsep{\fill}}lcccccc}
\toprule
\textbf{Method} & \textbf{Acc.} & \textbf{Prec.} & \textbf{Rec.} & \textbf{F1} & \textbf{FNR} & \textbf{FPR} \\
\midrule
BR-HIDF~\cite{he2023br}                    & 0.9767          & \textbf{0.9846} & 0.8924          & 0.9362          & 0.1076          & \textbf{0.0034} \\
Unit-Based~\cite{zhan2023adversarial}                 & 0.9876          & 0.9447          & \textbf{0.9946} & 0.9688          & \textbf{0.0054} & 0.0141          \\
Syscall-BSEM~\cite{zhang2021syscall}              & 0.9676          & 0.9659          & 0.8623          & 0.9106          & 0.1377          & 0.0075          \\
TfidfVectorizer~\cite{subba2021tfidfvectorizer}            & 0.9551          & 0.8662          & 0.9053          & 0.8853          & 0.0947          & 0.0331          \\
\textbf{Ours}              & \textbf{0.9948} & \textbf{0.9807}$_{2}$ & \textbf{0.9923}$_{2}$ & \textbf{0.9864} & \textbf{0.0077}$_{2}$ & \textbf{0.0046}$_{2}$ \\
\bottomrule
\end{tabular*}
\end{table}

\subsubsection{Ablation Study}
To analyze the contribution of each key component, we conduct ablation studies on the subject-level behavior aggregation module and the inter-subject relation modeling module. As shown in Table~\ref{tab:Ablation_comparison}, removing the subject-level behavior aggregation component (A1) leads to declines in Accuracy, Precision, Recall, and F1-score, with Precision and F1-score decreasing by 19.12 and 11.26 percentage points, respectively. As shown in Fig.~\ref{fig:roc_pr_comparison} (c) and (d), ROC-AUC and PR-AUC also decrease. These results indicate that relying solely on dispersed object-level operation features is insufficient to effectively integrate the behavioral information of the same subject across different objects, and that subject-level aggregation plays an important role in forming a complete behavioral representation. Removing the inter-subject relation modeling component (A2) similarly degrades detection performance, with Precision and F1-score decreasing by 17.34 and 11.08 percentage points, respectively, accompanied by reductions in ROC-AUC and PR-AUC. This result shows that even with subject-level representations, the absence of inter-subject runtime relationships weakens the model's ability to capture the overall semantics of malicious behavior.

Overall, these experiments verify the respective roles of subject-level behavior aggregation and inter-subject relation modeling in malicious behavior semantic learning. The results show that the hierarchical behavioral semantic learning method progressively integrates object-level operation features into subject-level and sample-level representations, enriching information across different levels and improving overall detection performance.

\begin{table}[htbp]
\centering
\caption{Performance Comparison of Ablation Studies}
\label{tab:Ablation_comparison}
\renewcommand{\arraystretch}{0.85}
\resizebox{\columnwidth}{!}{ 
\begin{tabular}{lcccccc}
\toprule
\textbf{Method} & \textbf{Acc.} & \textbf{Prec.} & \textbf{Rec.} & \textbf{F1} & \textbf{FNR} & \textbf{FPR} \\
\midrule
\makecell[l]{A1}    & 0.9464 & 0.7895 & 0.9783 & 0.8738 & 0.0217 & 0.0617 \\
\makecell[l]{A2} & 0.9485 & 0.8073 & 0.9565 & 0.8756 & 0.0435 & 0.0540 \\

\textbf{Ours}              & \textbf{0.9948} & \textbf{0.9807} & \textbf{0.9923} & \textbf{0.9864} & \textbf{0.0077} & \textbf{0.0046} \\
\bottomrule
\end{tabular}
}
\end{table}

\begin{figure}[!t]
    \centering
    \includegraphics[width=\columnwidth]{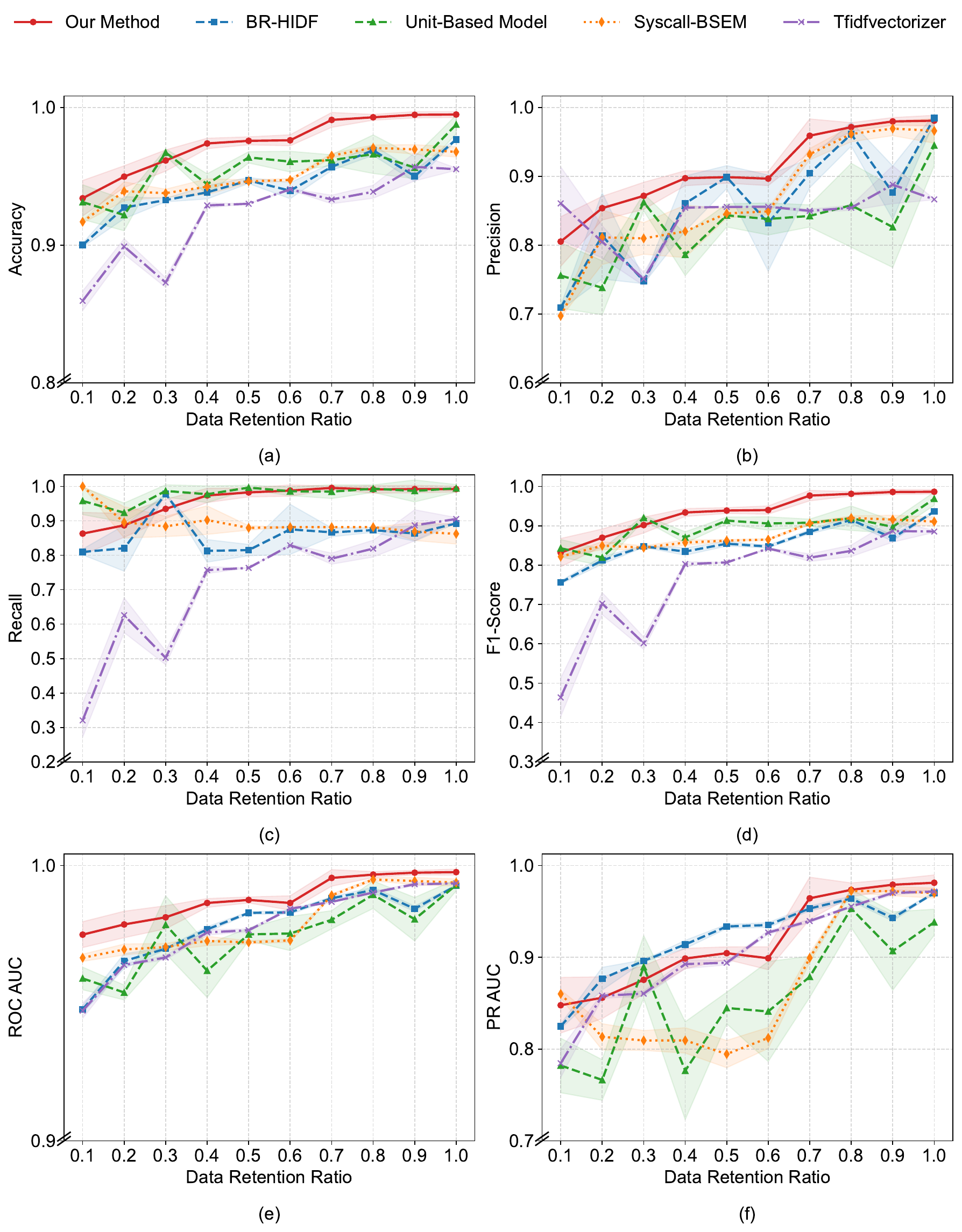}
    \caption{Performance comparison under different data retention ratios across six evaluation metrics.}
    \label{fig:model_performance_1x4}
\end{figure}

\subsubsection{Learning Curve Analysis}

To evaluate model performance under different amounts of training data, we vary the proportion of the training set and compare all models on the same test set, as shown in Fig.~\ref{fig:model_performance_1x4}. Overall, our method achieves the best or near-best results on most metrics, demonstrating stronger detection capability.
Specifically, when the training ratio ranges from 0.1 to 0.3, our method outperforms the baseline models on several key metrics, including Accuracy, Precision, and F1-score. However, its Recall is lower than that of the Unit-Based Model and remains at a moderate level overall. This may be because, with limited training samples, the model cannot yet fully learn the complex process collaboration patterns and deeper behavioral relationships involved in malicious activities, resulting in relatively more false negatives. In contrast, the Unit-Based Model remains highly sensitive to critical behavioral units and therefore maintains a higher Recall in the small-sample setting, although its overall classification stability and comprehensive performance remain limited. When the training ratio reaches 0.4, our method begins to show a clear overall advantage over the baseline models and takes the lead on several key metrics, including Accuracy, Precision, and F1-score. In the 0.7--1.0 range, even compared with the Unit-Based Model, our method generally maintains a comparable Recall while consistently outperforming all baseline models on core metrics such as F1-score, ROC-AUC, and PR-AUC. 
These results show that, as the amount of training data increases, our method can more effectively learn the key behavioral semantics of malicious activities than the other methods, thereby improving its overall detection capability.
In addition, across 10 independent runs at different training ratios, our method generally exhibits smaller standard deviations, whereas baseline methods such as BR-HIDF and the Unit-Based Model show more noticeable performance fluctuations. The proposed method also demonstrates a more stable and consistent improvement trend across different evaluation metrics, indicating that it can extract critical behavioral features of malicious activities more reliably across different training-set sizes and has stronger data utilization and generalization capabilities.

\section{Conclusion}

Existing system-call-based host intrusion detection methods are susceptible to behavioral semantic fragmentation caused by concurrent execution. To address this issue, we propose ReSHID, a resource-aware behavior reconstruction and hierarchical semantic learning framework for host intrusion detection. The proposed method reconstructs syscall sequences around individual objects to improve behavioral semantic continuity and the density of relevant information, and employs GATv2 to build an end-to-end model for extracting and learning coordination patterns among subjects. Experimental results demonstrate that our method achieves strong intrusion detection performance and stability. Our method can provide effective support for fine-grained HIDS in modern Linux environments, particularly in highly concurrent and containerized scenarios such as cloud computing platforms.

\section*{Acknowledgments}
The study was supported by the Key Laboratory of Data Protection and Intelligent Management, Ministry of Education, Sichuan University and also the Fundamental Research Funds for the Central Universities.

\bibliographystyle{IEEEtran}
\bibliography{ref}

\vfill

\end{document}